\documentclass[10pt,a4paper,twoside]{article}
\usepackage[utf8]{inputenc}
\usepackage[T1]{fontenc}
\usepackage[english]{babel}

\usepackage{geometry}
\usepackage[usenames,dvipsnames,svgnames,table]{xcolor}
\definecolor{citegreen}{rgb}{0.00,0.70,0.30}
\usepackage[colorlinks=true,
            linkcolor=blue,
            urlcolor=blue,
            citecolor=blue]{hyperref}

\usepackage{amsrefs}
\usepackage{amssymb}
\usepackage{amsmath}
\usepackage{amsthm}
\usepackage{mathrsfs}

\usepackage{xspace}

\usepackage{pgf}
\usepackage{tikz}
\usetikzlibrary{patterns}

\usetikzlibrary{arrows,automata}

\usetikzlibrary{calc}

\DeclareMathAlphabet{\mathpzc}{OT1}{pzc}{m}{it}

\usepackage{epsfig}
\usepackage{enumerate}
\usepackage{graphicx}

\usepackage[title]{appendix}

\makeatletter 
\renewcommand\@biblabel[1]{#1.} 
\makeatother

\numberwithin{equation}{section}

\newtheorem{theorem}{Theorem}

\newtheorem{prop}[theorem]{Proposition}

\newtheorem{lemma}{Lemma}[section]
\newtheorem{corollary}[theorem]{Corollary}
\newtheorem{remark}{Remark}[section]

\definecolor{redd}{rgb}{0.95,0.2,0.2}
\definecolor{gris}{rgb}{0.9,0.9,0.9}
\definecolor{greenn}{rgb}{0.1,0.6,0.2}

\definecolor{cmgray}{rgb}{0.7,0.7,0.7}

\definecolor{cmblue}{rgb}{0.2,0.5,0.8}

\providecommand{\B}{\mathbf}
\providecommand{\BS}[1]{\boldsymbol{#1}}

\providecommand{\D}{\mathbb}

\providecommand{\R}{\mathrm}

\newcommand{\eu}{\mathrm{e}}

\def\ttau{t}
\def\sone{s_1}
\def\stwo{s_2}
\def\qone{q_1}
\def\qtwo{q_2}

\def\diam{{\rm diam\,}}

\def\ball{\mathrm{B}}
\def\bball{\mathbf{B}}

\def\bf1{{\mathbf 1}}

\def\dist{\mathrm{dist}}

\def\card{{\rm card}\,}

\def\supp{{\rm supp\,}}

\def\one{\mathbf{1}}

\def\mes{{\rm mes\,}}
\def\Ker{{\rm Ker\,}}
\def\spann{{\rm Span\,}}

\def\rc{{\mathrm{c}}}

\def\lam{{\lambda}}

\def\eps{\epsilon}
\def\ffi{\varphi}

\def\esmp#1#2{\D{E}^{#1}\left[\, #2\, \right]}
\def\bigesm#1{\D{E}\big[\, #1\, \big]}

\def\lr#1{\langle#1\rangle}

\def\tBK{{ \widetilde{\BK} }}

\def\BN{\B{N}}

\def\BA{\mathbf{A}}
\def\BC{\mathbf{C}}
\def\BD{\mathbf{D}}

\def\BG{\mathbf{G}}
\def\BK{\mathbf{K}}

\def\BH{\mathbf{H}}

\def\BHni{\mathbf{H}^{\rm ni}}

\def\BGni{\mathbf{G}^{\rm ni}}

\def\BQni{\mathbf{Q}^{\rm ni}}

\def\BQ{\mathbf{Q}}
\def\BS{\mathbf{S}}
\def\BU{\mathbf{U}}

\def\BA{\mathbf{A}}
\def\BB{\mathbf{B}}

\def\BV{\mathbf{V}}

\def\Bx{\mathbf{x}}
\def\By{\mathbf{y}}

\def\hBw{\widehat{\mathbf{w}}}

\def\Ba{\mathbf{a}}
\def\Bb{\mathbf{b}}
\def\Bu{\mathbf{u}}
\def\Bv{\mathbf{v}}

\def\Bw{\mathbf{w}}
\def\Bz{\mathbf{z}}
\def\Bzero{\mathbf{0}}

\def\BDelta{{\boldsymbol{\Delta}}}

\def\BLam{\mathbf{\Lambda}}

\def\Ups{\Upsilon}
\def\tUps{\widetilde{\Upsilon}}
\def\hUps{\widehat{\Upsilon}}

\def\bp{\overline{p}}
\def\tzeta{\widetilde{\zeta}}

\def\tmu{\widetilde{\mu}}

\def\Const{{\rm{Const}}}

\def\DD{\D{D}}

\def\DP{\D{P}}
\def\tDP{\widetilde{\D{P}}}
\def\DR{\D{R}}

\def\DZ{\mathbb{Z}}
\def\DN{\D{N}}

\def\csM{\mathscr{M}}

\def\cH{\mathcal{H}}

\def\cM{\mathcal{M}}

\def\cR{\mathcal{R}}

\def\cZ{\mathcal{Z}}

\def\cR{{\mathcal{R}}}

\def\Id{{\R{Id\,}}}

\def\rd{{\R{d}}}
\def\rdH{{\R{d_H}}}
\def\rdS{{\R{d_S}}}

\def\rdH{{\mathrm{d}_{\mathcal H}}}

\def\rS{{\R{S}}}

\def\be{\begin{equation}}
\def\ee{\end{equation}}
\def\ba{\begin{array}{l}}
\def\ea{\end{array}}
\def\bal{\begin{aligned}}
\def\eal{\end{aligned}}

\def\ble{\begin{lemma}}
\def\ele{\end{lemma}}

\def\bthm{\begin{theorem}}
\def\ethm{\end{theorem}}

\def\bco{\begin{corollary}}
\def\eco{\end{corollary}}

\def\bre{\begin{remark}}
\def\ere{\end{remark}}

\def\btm{\begin{theorem}}
\def\etm{\end{theorem}}

\def\btm{\begin{theorem}}
\def\etm{\end{theorem}}

\def\bpr{\begin{prop}}
\def\epr{\end{prop}}

\def\ffi{{\varphi}}

\def\fB{\mathfrak{B}}
\def\fF{\mathfrak{F}}

\def\fS{\mathfrak{S}}

\def\om{{\omega}}
\def\Om{{\Omega}}

\def\eps{\epsilon}

\def\Lam{{\Lambda}}
\def\lam{{\lambda}}

\def\pr#1{\D{P}\left\{\,#1\,\right\}}
\def\esm#1{\D{E}\left[\, #1\, \right]}

\def\hesm#1{\widehat{\D{E}}\left[\, #1\, \right]}
\def\hesmI#1{\widehat{\D{E}}^{I}\left[\, #1\, \right]}
\def\bighesmI#1{\widehat{\D{E}}^{I}\big[\, #1\, \big]}

\def\pt{\partial}
\def\half{\frac{1}{2}}

\def\Uone{\textsf{(U1)}\xspace}

\def\Utwo{\textsf{(U2)}\xspace}
\def\Uthree{\textsf{(U3)}\xspace}

\def\Vone{\textsf{(V1)}\xspace}
\def\Vtwo{\textsf{(V2)}\xspace}
\def\Vthree{\textsf{(V3)}\xspace}

\def\Unif{{\rm Unif}}

\def\cond{\,\big|\,}

\begin{document}

\title{Comprehensive proofs of localization\\in Anderson models with interaction. I. \\Two-particle  localization estimates}


\author
{Victor Chulaevsky}

\date{}

\maketitle
\begin{abstract}
We discuss the techniques and results of the multi-particle Anderson localization theory
for disordered quantum systems with nontrivial interaction. After a detailed presentation of the approach
developed earlier by Aizenman and Warzel, we extend their results to the models with exponentially decaying, infinite-range
interaction.
\end{abstract}

\thispagestyle{empty}

\section{Introduction. The motivation and the model}  
\label{Sec:intro}

This manuscript is designed as the first issue in a mini-series aiming to present a survey of results and techniques
of the rigorous localization theory of disordered quantum systems with nontrivial interaction. The first results in this direction,
establishing the stability of Anderson localization in a two-particle system in $\DZ^d$ with respect to a short-range interaction
\cite{CS09a}, have been immediately followed by the proofs of exponential spectral localization (cf. \cite{AW09a,CS09b})
and exponential strong dynamical localization (cf. \cite{AW09a}) in $N$-particle systems, for any fixed $N\ge 2$.

As it often happens, the first proofs are not necessarily the shortest and simplest ones. In particular, the method
of \cite{CS09a,CS09b}, a multi-particle adaptation of the variable-energy Multi-Scale Analysis (VEMSA) was later replaced
by a significantly simpler one -- multi-particle fixed-energy MSA (MP FEMSA). Such a simplification had a price: one had to
design a spectral reduction (MP)FEMSA $\Rightarrow$ (MP)VEMSA; such a reduction was known for the systems without interaction,
and its extension to interactive systems required a special form of the eigenvalue concentration (EVC) estimates, technically
more involved than the conventional Wegner-type bounds. Another important ingredient of the spectral reduction came from
the toolbox developed  by Germinet and Klein \cite{GK01}, originally for the single-particle Anderson models.

The MPMSA, in its variable-energy version, was also extended to the continuous Anderson models \cite{CBS11,Sab13}.

On the other hand, there has been a considerable time interval between the pioneering work by Aizenman and Warzel
\cite{AW09a} on the MPFMM and the next bold step in that direction, made recently by Fauser and Warzel \cite{FW14} who
treated interactive Anderson models in $\DR^d$ with infinite-range interaction. Quite naturally, the spectral
analysis of unbounded random Schrodinger operators required various techniques which would be considerably simpler
in the case of lattice systems, where "hard" functional analysis is often reduced to elementary linear algebra.

Taking into account the complexity of the original works, it seems reasonable to present their key techniques and ideas
in a "nutshell", and in the simplest possible situation. This is how emerged the idea of the above mentioned the mini-series.

\vskip3mm

We start the first issue with a thorough presentation of the MPFMM, essentially for the reason that certain steps in
popularization of the MPMSA, in its variable- and fixed-energy variants, have been already made in a recent monograph
\cite{CS13}. This manuscript, however, is not limited to a mere review, for we show that the new ideas, developed by Aizenman and Warzel
\cite{AW09a}, can be easily adapted to the models with exponentially decaying interaction of infinite range.
The lattice models with infinite-range interaction were studied in our paper \cite{C12c} and, more recently, in our joint
work with Yuri Suhov \cite{CS14}. While a detailed presentation of the MPMSA is scheduled for subsequent issues, here
we discuss some similarities and particularities of the two approaches, MPFMM and MPMSA.

In the multi-particle models with a \emph{finite}-range interaction, the MPFMM, when
applicable, provides the strongest decay bounds upon the eigenfunction correlators (EFC),
as does its original, single-particle variant. In particular, such bounds are  stronger than those
proved with the help of the (single- or multi-particle) MSA, provided both methods apply to the same model.
However, the relations between the two approaches are more complex (at least, for now) in the realm of
multi-particle, interactive models than  for the systems with no interaction.

\vskip3mm

Traditionally, the FMM is an unchallenged
champion when it comes to the trees and other graphs with exponential growth of balls, while the MSA
demonstrates its unparalleled flexibility in the situations where the
"local"\footnote{We use this informal term, since, analytically, the description of disorder is not exactly the same
for the discrete and continuous Anderson models. In the case of the lattice systems with, say, an IID potential, the suitable
term here would be "single-site marginal" [distribution].}
probability distribution of the random
potential is not H\"{o}lder-continuous. As is well-known, for the lattice systems, it suffices to require the local distribution
to be at least log-H\"{o}lder continuous, and for the systems in $\DR^d$, with an alloy-type potential, the MSA establishes localization
for \textbf{\emph{any}} nontrivial probability distribution of the scatterers' amplitudes (cf. \cite{BK05,GK13}).

\vskip3mm

In the world of the interactive Anderson models, an additional parameter -- the decay rate of the interaction -- ap\-pears and, for the moment, sets apart the
results that can be (or rather, have been) proved by the MPFMM and the MPMSA; we discuss this issue below.

We focus on the discrete systems for the obvious reason that this requires a minimum of analytical tools, making
the proofs more comprehensive.

The choice of the \emph{two}-particle systems for this, first issue in the planned mini-series, is
motivated by several reasons.

$\bullet$ A number of geometrical  arguments become most simple
in the configuration space of $N=2$ particles, so the $2$-particle configuration space is $(\DZ^{d})^2\cong \DZ^{2d}$;
in the illustrations provided in the present paper, we refer to the case where $d=1$.

$\bullet$ It so happens that the existing decay bounds on the eigenfunctions (EFs) and on the
EF correlators (EFCs) have been first proven with respect not to a norm-distance in $(\DZ^d)^N$
(or, respectively, $(\DR^d)^N$), but to a pseudo-distance, used explicitly in \cite{AW09a}
and implicitly in \cite{CS09b}. Specifically, given two configurations of particles,
$\Bx$ with the particle positions $x_1, \ldots x_N$ and $\By$ with particles at
$y_1, \ldots y_N$,  this is the Hausdorff distance $\rdH(\Bx, \By)$ between the sets
of the respective particle positions. While this is
in general a complicated quantity, and a source of various unpleasant technical problems
in the $N$-particle localization analysis, in the particular case $N=2$, $\Bx = (x_1,x_2)$,
$\By = (y_1,y_2)\in(\DZ^d)^2$, one has the identity
$\rdH(\Bx,\By)=\rdS(\Bx,\By)$ where $\rdS(\cdot\,,\cdot)$ is the symmetrized
max-distance in $(\DZ^d)^2$:
$$
\rdS(\Bx,\By) = \min \big[ |\Bx - \By|_\infty, |\pi(\Bx) - \By|_\infty \big]
$$
with $|\Bz|_\infty := \max\big[ |z_i|, 1 \le i \le d \big]$ and (the only nontrivial) permutation $\pi\in \rS_2$
exchanges the particle positions: $\pi(x_1,x_2) = (x_2,x_1)$.

More to the point, the symmetrized max-distance is the natural max-distance in the configuration space
of a system of two indistinguishable particles in $\DZ^1$. The latter can be
implemented\footnote{For $d\ge 1$, the construction is slightly more complicated.}
as the
"half-space" $\{\Bx\in \DZ^2:\, (x_1 < x_2)\}$ (Fermi-particles) or, respectively,
$\{\Bx\in \DZ^2:\, (x_1 \le x_2)\}$ (Bose-particles).

$\bullet$ The most significant reason for choosing $N=2$ comes from the regrettable fact that,
although the above mentioned difficulty, appearing for $N\ge 3$ particles, had been partially overcome in \cite{C10a,C12b,C13a},
the solution proposed there applies so far to a limited class of random potentials. On the bright side, this class
contains in particular the two most popular models of disorder used in physics (Gaussian distribution
and uniform distribution in a finite interval), yet one is still far from the wealth of rigorous mathematical results
of the $1$-particle localization theory, where in the
lattice
systems of dimension $>1$ it suffices
to require the probability distribution function (PDF) of the random potential to be log-H\"{o}lder continuous.

\vskip3mm

Summarizing, the $2$-particle systems have been selected for this first issue in order
to present the original methods from \cite{AW09a} and \cite{CS09a} in their best possible light, and with
a minimum of technicalities that can easily start obscuring the key ideas in more general models.

\vskip3mm

$\blacktriangleright$ The mathematical theory of Anderson models with interaction is actually full of surprises. One of them is that,
in contrast to the conventional, $1$-particle theory, where since 1993 one has had two
alternative, and mutually complementing, methods -- one based on a multi-scale geometrical induction (Multi-Scale Analysis = MSA,
going back to the pioneering works \cite{FS83,FMSS85}), and the other using, in a manner of speaking, a ``mono-scale'' strategy
(the Aizenman--Molchanov =AM method, further developed in a series of subsequent works
bearing a distinctive mark of Michael Aizenman's enthusiasm, cf., e.g., \cite{Ai94,ASFH01,AENSS06}),
these two approaches in the multi-particle localization theory
finally settle on the common ground of the \emph{multi-scale induction}, although they keep their distinctive features:
the multi-particle MSA (MPMSA) makes use of bounds in probability, while its counterpart (MPFMM) developed by Aizenman and Warzel
continues to successfully employ bounds in expectation.

$\blacktriangleright$ Another package of surprises (at least, that's the way it looks so far) awaits one when it comes to the analysis of localization
in presence of interactions of infinite range, decaying slower than exponentially. First, the MPFMM, applied recently by
Fauser and Warzel \cite{FW14} to the interacting systems in $\DR^d$, no longer provides \emph{exponential} bounds on the EFCs
and, as a result, on the decay of the eigenfunctions. Secondly, the MPMSA is no longer at a disadvantage (compared to the MPFMM)
when it comes to the decay rate of the EFCs, i.e., the rate of the Strong Dynamical Localization (=SDL), and even provides
the  strongest possible result -- an exponential decay -- relative to the \emph{eigenfunctions}.

$\blacktriangleright$ Finally, recall that we have already mentioned yet another surprise of the multi-particle Anderson theory:
the physically most sound
decay bounds (on EFs and on EFCs) for $N\ge 3$ particles, viz. the bounds in a (symmetrized) norm-distance and not the Hausdorff distance,
have been established so far only with the help of the MPMSA.

It is to be stressed that the present manuscript is most certainly not intended as a replacement for
(but only a complement to)
the exposition in \cite{AW09a}. For instance, Aizenman and Warzel
address in \cite{AW09a} the problem of perturbative parametric stability of the Anderson localization
under a sufficiently weak interaction, in the situation where the $1$-particle system exhibits
strong localization in terms of the fractional moments. The most notable example of such a situation
is localization of weakly (and locally) interacting one-dimensional quantum particles in a random environment.

\subsection{The multi-particle Hamiltonian}\label{ssect:LaplHsam} 

For the reasons explained above, we consider the random Hamiltonian $\BH(\om)$ acting as
a bounded self-adjoint operator in $\ell^2((\DZ^d)^2)$, of the form
$$
\BH(\om) = \sum_{j=1}^2 \Big( H_0^{(j)} + g V(x_j;\om)  \Big) + \BU(\Bx)
$$
where $V:\DZ^d\times\Om\to\DR$ is a random field on $\DZ^d$ (the configuration space of single particles),
relative to a probability space $(\Om,\fF,\DP)$, $\BU$ is the operator of multiplication by the interaction potential
$(x_1,x_2)\mapsto U(|x_1-x_2|)$,
and $H_0^{(j)}$, $j=1,2$, are replicas of the standard, nearest-neighbor lattice Laplacian on $\DZ^d$, acting
on the respective variables (particle positions) $x_j$.

\subsection{Assumptions}

In Theorems \ref{thm:Main.finite}, \ref{thm:Main.infinite} and \ref{thm:MPMSA} we consider the interaction
potentials satisfying one the following conditions.

\vskip2mm
\Uone $\supp U \subset [0,r_0]$, $r_0 < +\infty$.

\vskip2mm
\Utwo $U(r) \le \eu^{-m'r}$, $m'>0$.

\vskip2mm
\Uthree $U(r) \le \eu^{-m'r^\zeta}$, $m', \zeta>0$.

\vskip2mm
\noindent
Our main assumption on the external (random) potential is as follows.

\vskip2mm
\Vone  The random field $V:\DZ^d\times\Om\to\DR$ is IID, a.s. bounded, with
$$
\pr{ V(x;\om) \in [0,1] }=1,
$$
and admits a bounded (common) marginal probability density $p_V$, with $\| p_V \|_\infty = p_V <+\infty$.

\vskip2mm

The non-negativity of $V$ is, of course, inessential, for the transformation $V \mapsto V + E$ results only in a spectral
shift for $\BH$, leaving its eigenfunctions (of whatever nature) invariant. Making larger the amplitude of $V$
amounts to taking larger $|g|$.

\vskip3mm
The situation with the hypotheses required for the proof of Theorem  \ref{thm:MPMSA} is a bit more complicated.

Its two-particle version, stated in this paper, does not actually require the existence of a bounded density; neither does
the technique from \cite{AW09a}. In both cases, one can relax the condition of Lipschitz continuity of the marginal
distribution to that of H\"{o}lder continuity.

On the other hand, in the general case where $N>2$, the conditions depend at the moment upon the desired result:
decay estimates in the Hausdorff distance still can be proved for $V$ with bounded density (or even H\"{o}lder-continuous PDF),
but a more suitable decay in a norm-distance in the $N$-particle configuration space $(\DZ^d)^N$
requires the following, stronger condition upon the random field $V$. As was already said, norm-distance
bounds have not yet been proved with the help of the MPFMM.

Ref. \cite{CS14}, where  localization bounds in a norm-distance were proved, relies on the following assumption
(required \emph{only} for $N>2$):

\vskip2mm
\Vtwo  The random field $V:\DZ\times\Om\to\DR$ is IID, a.s. bounded, with
$$
\pr{ V(x;\om) \in [0,1] }=1,
$$
and admits a bounded marginal probability density $p_V$, with
$$
 p_*\one_{(0,1)} \le p_V(t)\one_{(0,1)} \le \bp\one_{(0,1)},
$$
$0 < p_* < \bp < +\infty$, and bounded derivative $p'_V$ on $(0,1)$.

\vskip2mm

The last condition certainly looks strange to a reader familiar with the eigenvalue concentration bounds
(starting with the celebrated Wegner bound) for random Anderson Hamiltonians. We will explain the reasons
for this condition in a forthcoming manuscript; here we simply mention that such a hypothesis appears
in Ref.~\cite{C13a} used in \cite{CS14}.

In the case $N=2$, the  restrictive hypothesis \Vtwo can be substantially relaxed;
the main assumption in Theorem \ref{thm:MPMSA} is as follows:
\vskip2mm
\Vthree  The random field $V:\DZ\times\Om\to\DR$ is IID, with uniformly H\"{o}lder continuous marginal probability distribution function (PDF)
$F_V(t) := \pr{ V(x;\om) \le t}$: for all $s\in [0,1]$ and some $b\in(0,1]$, $C<\infty$
$$
 \sup_{t\in\DR} \big( F_V(t+s) - F_V(t) \big) \le C s^b .
$$

\vskip2mm

The disorder amplitude $g>0$, which we assume in this paper to be large enough, can be introduced, e.g., by putting a small
factor $g^{-1}$ in front of the kinetic operator or a large factor $g$ in $V$, or else in a more subtle way, by assuming
the marginal probability density of the IID random field $V$ to be small, viz. of order of $O(g^{-1})$. For definiteness,
and also in order to follow more closely \cite{AM93} and \cite{AW09a}, we consider the potential of the form
$gV(x;\om)$ with $g \gg 1$. On one occasion (see Sect.~\ref{sec:finiteness.FM}), it will be convenient to change the attitude
and work with operators $\BV + g^{-1}\BA$, where $\BA$ is the extended kinetic operator $\BH_0 + \BU$.

As was pointed out in Ref.~\cite{AW09a}, the assumption of existence and boundedness of the single-site marginal
density of the random potential can be relaxed to H\"{o}lder continuity. However, it seems that this would require a modification
of the proof of the uniform a priori bound on the fractional moments in  Sect.~\ref{sec:finiteness.FM}. Specifically, the standard argument
employed in the proof of the so-called weal $L^1$-bound, based on the linear transformation of the two-dimensional space
supporting the reduced probability measure, requires a slightly more elaborate approach, developed in Ref.~\cite{AM93}.

\section{Basic notation and preliminary remarks}

In the context of multi-particle Hamiltonians, we usually employ boldface
notation for the objects referring to a multi-particle system, to visually distinguish them from their single-particle counterparts.

Recall the definition of the Hausdorff distance.

Given two subsets $X,Y$ of an abstract metric space $(\csM,\rd)$, the Hausdorff distance between these
subsets is given by
$$
\rdH(X,Y) := \max \Big[\;  \sup_{x\in X} \rd(x, Y), \; \sup_{y\in Y} \rd(y, X)  \; \Big] .
$$

Associating with a configuration of $N$ distinguishable particles in $\DZ^d$,
$\Bz = (z_1, \ldots, z_N)$, its "projection" to the $1$-particle
configuration space $\DZ^d$, $\Pi \Bz = \{z_1\} \cup \cdots \{z_N\}$, we extend $\rdH$, defined for the \emph{subsets}
of $\DZ^d$, to the particle configurations:
$\rdH(\Bx,\By) := \rdH(\Pi \Bx, \Pi\By)$.

The following elementary statement explains what makes the $2$-particle systems
so special in the framework of our analysis.

\ble[\emph{Hausdorff distance equals max-distance in $(\DZ^d)^2$}]
\be\label{eq:dH.eq.dS}
\forall\, \Bx,\By\in\DZ^2\quad \rdH(\Bx, \By) = \rdS(\Bx, \By).
\ee
\ele
\proof
Let $R := \rdS(\Bx, \By) = \min_{\pi\in S_2} |\pi(\Bx) - \By|$. Then for at least one choice of the vectors
$\Ba=(a,a')\in\{ (x_1,x_2), (x_2,x_1)\}$ and $\Bb=(b,b')\in\{ (y_1,y_2), (y_2,y_1)\}$,
with  $\{a,a'\} = \Pi \Bx$, $\{b,b'\} = \Pi \By$,
one has
$$
\begin{array}{lll}
&\left.
\begin{array}{lc}
| a - b| &= R
\\
| a' - b'| & \le  R
\end{array} \right\} & \text{ i.e., } |\Ba - \Bb| = R,
\\
&
\max \big[ | a - b'|, \; | a' - b| \big]  \ge R,
&  \text{ i.e., } |\pi(\Ba) - \Bb| \ge R \text{ for  } \Id \ne \pi\in S_2.
\end{array}
$$

If $| a - b'| \ge R$, then $\rd(a, \Pi \Bb) = R$, while
$\rd(a', \Pi \Bb) \le \rd(a', b')\le R$. In this case, we conclude that
$\rdH(\Bx,\By) \equiv \rdH(\Ba, \Bb) = R = \rdS(\Bx,\By)$.

In the remaining case where $|a' - b| \ge R$, the argument is similar
and the final conclusion is the same.
\qedhere

\vskip2mm

From this point on, we use only $\rdS$, even in the arguments where $\rdH$ would be required in the case $N\ge 3$.

For a proper subset $\varnothing\ne \Lam\subsetneq \DZ^d$,
we denote $\pt^-\Lam =\{y\in\Lam:\, \dist(x,\Lam^\rc=1 \}$, $\pt^+\Lam = \pt^- \Lam^\rc$, with $\Lam^\rc := \DZ^d\setminus\Lam$.

We will systematically make use of the elementary inequality which is one of the cornerstones of the
FMM technique: $\forall\, s\in(0,1)$  $\big| \sum_n a_n \big|^s \le \sum_n |a_n|^s$.

Given finite lattice subsets $\BLam_1 \subsetneq \BLam\subset (\DZ^{d})^2$, with the edge boundary of $\BLam_1$
relative to $\BLam$,
$$
\pt \BLam_1 = \pt^{(\BLam)} \BLam_1 := \{(\Bx,\By)\in \BLam_1\times \BLam^\rc: \, |\Bx-\By|=1\}
$$
one can easily infer from the second resolvent identity the so-called
Geometric Resolvent Equation (GRE) and the
Geometric Resolvent Inequality (GRI) (cf., e.g., \cite{Kir08b}). The latter will be used in its FMM-flavoured form (which we will call the FGRI):
for any $s\in(0,1)$, $\Bx\in\BLam_1$, $\By\in\BLam\setminus \BLam_1$, and for any $E$
such that $(\BH_{\BLam} - E)$ and $(\BH_{\BLam_1} - E)$ are invertible,
\be\label{eq:GRI.1}
|\BG_\BLam(\Bx,\By;E)|^s \le \sum_{(\Bw,\Bw')\in \pt \BLam_1} |\BG_{\BLam_1}(\Bx,\Bw;E)|^s\,  |\BG_\BLam(\Bw',\By;E)|^s.
\ee

A detailed discussion of various boundary conditions for the LSO (and of the related forms of the GRI)
can be found in the review by Kirsch \cite{Kir08b}.

For brevity and folowing \cite{AW09a}, we introduce the energy-disorder expectation $\hesmI{\cdot}$: given
a measurable function $f: \DR \times \Om$ and an interval $I\subset \DR$, we set
$$
\hesmI{ f(E,\om) } := |I|^{-1} \int_I \esm{ f(E,\om) } \, dE,
$$
where $\esm{\cdot}$ is the conventional expectation relative to $(\Om,\DP)$. For brevity, we keep the superscript "I"
only where necessary or instructive and often write $\hesm{\cdot}$ instead of $\hesmI{\cdot}$. Since the measure
$|I|^{-1} \, dE$ on $I\subset \DR$ is normalized, the inequality \eqref{eq:GRI.1} implies a similar bound for the  expectation
$\hesm{\cdot}$ relative to the augmented probability space $I \times \Om$.

\subsection{Structure of the paper}

The central result presented in this paper is Theorem \ref{thm:Main.finite},
proved by Aizenman and Warzel \cite{AW09a}, while a more general Theorem \ref{thm:Main.infinite}
follows by a relatively simple adaptation of just one important ingredient of the proof (cf. Lemma  \ref{lem:exp.dH.diamx.diamy.2}
in Sect.~\ref{sec:split.infinite}).

We keep the main flow of argument in the proof of  Theorem \ref{thm:Main.finite} as
``linear'' as possible, yet two  exceptions seem appropriate.

$\bullet$ The first one concerns a key component of the FMM -- an a priori bound by $O(1)$ (indeed, by $O(|g|^{-s}$) on
the fractional moments (cf. Sect.~\ref{sec:finiteness.FM}).

$\bullet$ The second one (cf. Sect.~\ref{sec:split.finite})
is a fairly simple proof of decay bounds away from the support of the interaction $\BU$; see Fig. 1.

The proofs of these two results can be read independently of the main body of the proof; the latter
is essentially a \emph{multi-scale induction}.

Appendix \ref{app:Boole} could have been replaced with a mere reference to the work by Boole \cite{Boole}, but we break here with this
common practice for several reasons. Firstly, Ref. \cite{Boole} is not quite easy to find in libraries, although,
with a bit of effort,
it can be found in a downloadable form on Internet. Secondly, it would require at least an adaptation,
both stylistic (recall that \cite{Boole} was written in 1857) and notational. Thirdly, it seems that the shortest
proof was given by Loomis \cite{Loo46}, whose argument we reproduce almost verbatim in Appendix \ref{app:Boole}.

\section{Main results on the decay of the GFs}

\btm[Following Aizenman and Warzel \cite{AW09a}]\label{thm:Main.finite}
Assume that the interaction potential $U$ is compactly supported
(cf. Assumption {\rm \Uone})  and the external random potential satisfies Assumption {\rm \Vone}.
There exists $g_0\in(0,+\infty)$ with the following properties.

For any finite connected subgraph $\Lam\subset\DZ^d$ and for all $g$ with $|g|\ge g_0$, the two-particle Hamiltonian $\BH_{\Lam^2}(\om)$
exhibits exponential decay of the fractional moments
of the Green functions. Specifically, for some finite interval $I\subset \DR$ containing the a.s. spectrum
of $\BH_{\Lam^2}(\om)$, one has, with $m = m(g) \to+\infty$ as $g\to+\infty$,
\be
\bighesmI{ |\BG_{\Lam^2}(\Bx,\By;E)|^s}  \,dE  \equiv \int_I \bigesm{ |\BG_{\Lam^2}(\Bx,\By;E)|^s}  \,dE  \le \eu^{ - m \rdS(\Bx,\By)} .
\ee
The same upper bound holds true for the two-particle Hamiltonian in the entire lattice $\DZ^d$, i.e., for
$\BH_{(\DZ^d)^2}(\om)$.
\etm

The result of Theorem \ref{thm:Main.finite} implies in fact strong dynamical localization
(hence, spectral localization with probability one).
Such an implication is neither new (this was done in \cite{AW09a}) nor difficult to prove, with the help of
technical tools developed by Elgart et al. \cite{ETV10} (see also \cite{CS13}) and by Germinet and Klein
\cite{GK01}. We plan to address it in a forthcoming issue.
Recall that Ref.~\cite{AW09a} provides a complete proof of exponential strong dynamical and spectral localization in the situation
more general than the one covered by Theorem \ref{thm:Main.finite}.

Using the augmented, energy-disorder measure space $I\times\Om$ is mainly motivated by Lemma \ref{lem:EFC.to.GF}
(cf. \cite{AW09a}*{Theorem 4.2}), allowing one
to transform the decay properties of the EFCs into those of the [integrated] Green functions. On the other hand,
it seems appropriate to stress here that
the energy-disorder space $I\times\Om$ was used already at an early stage of the development of the MSA by
Martinelli and Scoppola \cite{MS85} who proved that fast decay of the Green functions implied absence of a.c. spectrum;
Refs. \cite{ETV10} and \cite{CS13} improve this result.

We will focus on the finite-domain Hamiltonians $\BH_{\Lam^2}(\om)$, since the extension to an infinitely extended domain
is obtained by a simple application of the Fatou lemma on converging measures. This argument can be found in
Refs.~\cites{ASFH01,AENSS06}; it is not specific to the localization analysis carried out with the help of the FMM.

Once the proof of Theorem \ref{thm:Main.finite} (for finite $\Lam\subset\DZ^d$)
is completed, we will show that the main approach extends with no
difficulty to infinite-range, exponentially decaying interactions, thus generalizing the original results
by Aizenman and Warzel \cite{AW09a}.

\btm\label{thm:Main.infinite}
Assume that the interaction potential $U$ in the two-particle Hamiltonian $\BH(\om)$ decays exponentially fast at infinity
(cf. Assumption {\rm \Utwo}) and the external random potential satisfies Assumption {\rm \Vone}.
For $g_0$ large enough and for all $g$ with $|g|\ge g_0$, $\BH(\om)$ exhibits exponential decay of the fractional moments
of the Green functions:
\be
\int_I  \bigesm{ |\BG(\Bx,\By;E)|^s} \,dE \le \eu^{ - m \rdS(\Bx,\By)}
\ee
for some finite interval $I\subset \DR$ containing the a.s. spectrum of $\BH(\om)$.
\etm

Here a similar remark can be made, concerning the derivation of spectral and dynamical localization from the decay of the Green functions.

Our proof is closer to the original technique from \cite{AW09a} than to a more advanced method developed by
Fauser and Warzel \cite{FW14} for differential operators.

Note also that Theorem \ref{thm:Main.infinite} naturally and easily extends to more general locally finite graphs $\cZ$,
replacing $\DZ$, satisfying a condition of polynomially bounded growth of balls. We plan to discuss such an extension
in a forthcoming issue.

On the other hand, the approach of Ref.~\cite{FW14} gives rise only to a sub-exponential decay of eigenfunction correlators,
when $U$ decays slower than exponentially, and this results to a sub-exponential decay bound on the eigenfunctions.
As was said in the Introduction, here the MPFMM and the MPMSA are on essentially equal footage,
and the latter even provides stronger (exponential) decay bounds on the eigenfunctions (albeit not on the EFCs).
For this reason, we postpone to a forthcoming issue a detailed presentation of the MPFMM techniques in the case of sub-exponentially
decaying interactions.

\subsection{Statement of results on exponential decay of eigenfunctions}

Here we briefly describe the kind of decay estimates that can be proved for the $2$-particle
systems with sub-exponentially decaying interaction, using the multi-particle multi-scale analysis.

\btm[Cf. \cite{CS14}]\label{thm:MPMSA}
Assume that the interaction potential $U$ in the two-particle Hamiltonian $\BH(\om)$  satisfies Assumption {\rm \Uthree},
and the external potential satisfies {\rm \Vthree}.
Then for $g_0>0$ large enough and for $|g|\ge g_0$,
with probability one, the two-particle Hamiltonian has pure point spectrum, and
all its eigenfunctions decay \textbf{exponentially fast} at infinity.
The eigenfunction correlators decay sub-exponentially fast at infinity.
\etm

As was already said, in the more general case where $N\ge 3$, Assumption \Vthree is to be replaced by a more restrictive
one, \Vtwo; otherwise, the existing techniques give rise only to Hausdorff-distance (and not norm-distance) decay of the EFCs.


\section{Proof of Theorem \ref{thm:Main.finite}}
\label{sec:proof.Main.finite}

We follow closely the arguments from \cite{AW09a} (without repeating it every time again),
taking shortcuts thanks to the simplifying assumption $N=2$, and adapting notation and calculations on an as-needed basis.

We will use the sequence of length scales $\{L_k, \, k\ge 0\}$ defined by the recursion
\be\label{eq:def.Lk}
L_{k+1} :=  2(L_k + 1), \;\; k=0, 1, \ldots,
\ee
or, explicitly,
\be\label{eq:def.Lk.explicit}
L_{k} =  2^k(L_0 + 2) - 2.
\ee
It suffices to treat the case where $\Bx,\By$ with $\rdS(\Bx,\By) =: R>L_0$, for otherwise the decay bound can be absorbed
in the constant factor. Given $\Bx,\By$ with $\rdS(\Bx,\By) =: R>L_0$, there is a unique integer $k\ge 0$ such that
\be\label{eq:dH.Lk.L.k+1}
 L_k < R = \rdS(\Bx,\By) \le L_{k+1}.
\ee
This value $k$ will be fixed until the end of the proof. Since $k$ can be arbitrarily large, the proof will require a scale induction.

Further, assume w.l.o.g. that $\diam \Pi\Bx \ge \diam \Pi\By$.

Next, by definition of the max-distance,
there are the points $x\in\Pi \Bx$, $y\in\Pi \By$, such that
\be\label{eq:dH.x-y}
\rdS(\Bx,\By) = R = |x-y|.
\ee
By the stochastic translation invariance of the random field $V$, and the resulting diagonal shift-invariance of the EF correlators,
we can assume w.l.o.g. that $x=0$; this reduction is made mainly for making simpler the representation on Fig. 1,
where it is instructive to indicate the positions of both coordinate axes (the replicas of the physical, $1$-particle
configuration space, assumed to be $\DZ^1$). With this reduction, for the rest of the proof, we have that
\begin{enumerate}[(i)]
  \item $\Pi \Bx \ni 0$;
  \item there is a point $y\in\Pi \By$ such that
\be\label{eq:R.dH.x-y}
\bal
   |y| \equiv |y-0| = \rdS(\Bx, \By) = R > L_k.
\eal
\ee
\end{enumerate}

%
%
\begin{tabular}{lll}
\begin{tikzpicture}
\begin{scope}[scale=0.50]
\clip (-14.5,-12.0) rectangle ++(23.0,20.5);

\begin{scope}
\clip (-7.4,-7.4) rectangle ++(15.3,15.3);
\draw[color=gray!50!white!40, line width = 58] (-7.5,-7.5) -- (7.9,7.9);

\draw[color=lightgray, line width = 17] (-7.5,-7.5) -- (7.9, 7.9);

\end{scope}

\foreach \i in {-7, -6.5, -6, -5.5, -5, -4.5, -4, -3.5, -3, -2.5, -2, -1.5, -1, -0.5, 0, 0.5, 1, 1.5, 2, 2.5, 3, 3.5, 4, 4.5, 5, 5.5, 6, 6.5, 7, 7.5  }
\draw[color=gray, line width = 0.5] (\i, -7.4) -- (\i, 7.9);

\foreach \i in {-7, -6.5, -6, -5.5, -5, -4.5, -4, -3.5, -3, -2.5, -2, -1.5, -1, -0.5, 0, 0.5, 1, 1.5, 2, 2.5, 3, 3.5, 4, 4.5, 5, 5.5, 6, 6.5, 7, 7.5  }
\draw[color=gray, line width = 0.5] (-7.4, \i) -- (7.9, \i);

\draw[color=black,, line width = 0.7, pattern=north west lines, pattern color=gray] (-1, -1) rectangle ++(2.0, 2.0);

\draw[color=black,, line width = 0.7] (-2, -2) rectangle ++(4.0, 4.0);

\draw[color=black!50!white, line width = 2.7] (-4.5, -4.5) rectangle ++(9.0, 9.0);

\draw[color=black!50!white, line width = 2.7] (-4.5, -4.5) rectangle ++(9.0, 9.0);

\draw[color=black!100!white, line width = 1.2] (-5, -5) rectangle ++(10.0, 10.0);

\draw[color=black!30!white, line width = 4] (2.5, 4.5) -- (4.5, 4.5);
\draw[color=black!30!white, line width = 4] (4.5, 4.5) -- (4.5, 2.5);

\draw[color=black!30!white, line width = 4] (-2.5, -4.5) -- (-4.5, -4.5);
\draw[color=black!30!white, line width = 4] (-4.5, -4.5) -- (-4.5, -2.5);

\foreach \i in {-2.5, -3, -3.5, -4, -4.5, -5, -5.5, -6, -6.5}
\fill[color=gray] (\i, -4.5) circle  (0.15);

\foreach \i in {-2.5, -3, -3.5, -4, -4.5, -5, -5.5, -6, -6.5}
\fill[color=gray] (-4.5, \i) circle  (0.15);

\foreach \i in {2.5, 3, 3.5, 4, 4.5}
\fill[color=gray] (\i, 4.5) circle  (0.15);

\foreach \i in {2.5, 3, 3.5, 4, 4.5, 5, 5.5, 6, 6.5}
\fill[color=gray] (4.5, \i) circle  (0.15);

\foreach \i in {-7, -6.5, -6, -5.5, -5, -4.5, -4, -3.5, -3, -2.5, -2, -1.5, -1, -0.5, 0, 0.5, 1, 1.5, 2, 7, 7.5}
\fill[color=black] (4.5, \i) circle  (0.15);

\fill[color=gray!40!white, line width = 1.2] (4.5, 3) circle  (0.12);
\draw[color=blue, line width = 1.2] (4.5, 3) circle  (0.15);

\foreach \i in {-7, -6.5, -6, -5.5, -5, -4.5, -4, -3.5, -3, -2.5, -2, -1.5, -1, -0.5, 0, 0.5, 1, 1.5, 2}
\fill[color=black] (\i, 4.5) circle  (0.15);

\foreach \i in {-7, -2, -1.5, -1, -0.5, 0, 0.5, 1, 1.5, 2, 2.5, 3, 3.5, 4, 4.5, 5, 5.5, 6, 6.5, 7, 7.5}
\fill[color=black] (\i, -4.5) circle  (0.15);
\foreach \i in {-7, -2, -1.5, -1, -0.5, 0, 0.5, 1, 1.5, 2, 2.5, 3, 3.5, 4, 4.5, 5, 5.5, 6, 6.5, 7, 7.5}
\fill[color=black] (-4.5, \i) circle  (0.15);

\foreach \i in {-7, -6.5, -6, -5.5, -5, -4, -3.5, -3, -2.5, -2, -1.5, -1, -0.5, 0, 0.5, 1, 1.5, 2, 2.5, 3, 3.5, 4, 4.5, 5, 5.5, 6, 6.5, 7, 7.5 }
\fill (\i, \i) circle  (0.045);

\draw[color=black, line width = 1.2] (-7.5,0) -- (7.9, 0);
\draw[color=black, line width = 1.2] (0,  -7.5) -- (0, 7.9);

\draw[pattern=north west lines, pattern color=gray!70!white] (0, -10) circle (0.95);

\draw[pattern=north west lines, pattern color=gray!70!white] (2, -10) circle (0.95);

\draw[color=black, line width = 1.2] (-7.5,-10) -- (7.9, -10);

\fill[color=blue] (0.0, -0.5) circle  (0.15);

\fill[color=green!70!blue] (2.0, -1) circle  (0.10);
\fill[color=gray] (2.0, 0.5) circle  (0.10);

\draw[color=gray, line width =1] (2.5, 0.5) circle  (0.10);

\draw[color=green!70!blue, line width = 1] (2.5, -1) circle  (0.10);

\fill[color=black] (2.0, -10) circle  (0.10);

\fill[color=black] (4.5, -10) circle  (0.15);
\fill[color=black] (0.0, -10) circle  (0.15);

\draw[<->,color=black, line width = 1.2] (0.05,-11.5) -- (4.5, -11.5);
\node at (2.5, -11.2) (R) {$R$};

\node at (0.75, -1.7) (Bx) {$\Bx$};
\draw[->,bend right=30, line width = 0.7] (Bx) to (0.2, -0.55);

\node at (-0.75, -8.7) (x) {$x=0$};
\draw[->,bend right=30, line width = 0.7] (x.south) to (-0.2, -9.8);
\draw[dotted,color=black, line width = 1] (0.0,-7.5) -- (0.0,-10.0);

\node at (5.75, -8.7) (y) {$y$};
\draw[->,bend left=30, line width = 0.7] (y.south) to (4.65, -9.8);
\draw[dotted,color=black, line width = 1] (4.5,-7.5) -- (4.5,-10.0);

\node at (3.75, -8.7) (w) {$w$};
\draw[->,bend left=30, line width = 0.7] (w.south) to (2.1, -9.8);
\draw[dotted,color=black, line width = 1] (2.0,-7.5) -- (2.0,-10.0);

\node at (1.7, -2.7) (Bw) {$\Bw$};
\draw[->,bend left=30, line width = 0.7] (Bw) to (1.9, -1.1);
\draw[->,bend left=30, line width = 0.7] (Bw) to (1.85, 0.45);

\node at (2.9, -2.7) (Bwp) {$\Bw'$};
\draw[->,bend right=30, line width = 0.7] (Bwp) to (2.55, -1.15);
\draw[->,bend right=30, line width = 0.7] (Bwp) to (2.6, 0.4);

\node at (6.5, 2.7) (By) {$\By$};
\draw[->,bend left=20, line width = 0.7] (By) to (4.7, 2.9);

\end{scope}
\end{tikzpicture}
\\

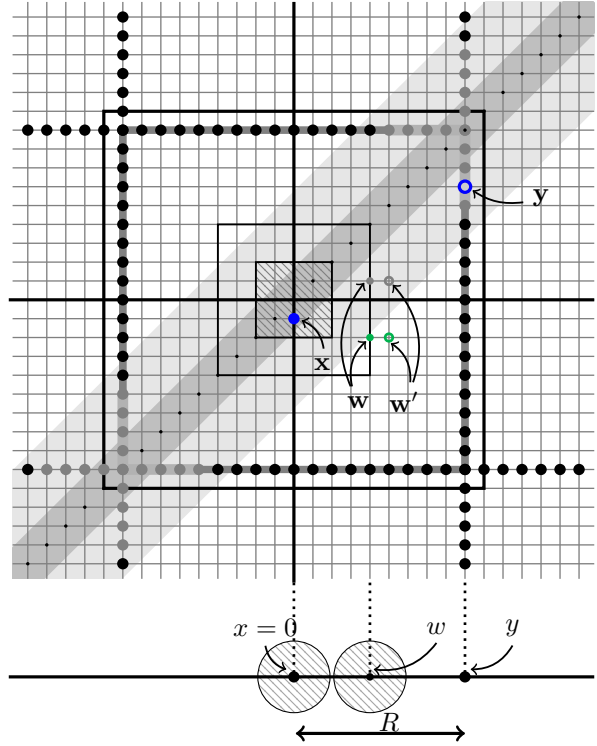
\centerline{\textbf{Figure 1.} Example for the proof of Theorem 1, with $d=1$, $N=2$.}
\end{tabular}
\vskip3mm

We fix an arbitrarily large but finite connected subset $\Lam\subset\DZ^d$; the connectedness is understood in the sense
of graphs: $\DZ^d$ is endowed with the graph structure where a pair $(x,y)$ is an edge iff $|x-y|_1 := \sum_i |x_i - y_i|=1$,
and $\Lam$ is considered as a subgraph of $\DZ^d$. For a given finite $\Lam$, the scale induction described below is to be stopped, once the scale
$L_k$ achieved at the $k$-th induction step is bigger or equal to $\diam \Lam$.

The finiteness of $\Lam$ allows us to work
with well-defined, finite-dimensional self-adjoint operators and their resolvents; the corresponding spectra are finite subsets of $\DR$.

The fractional moments $\esm{ |\cdot|^s}$ figuring in our formulae are computed  for $s\in(0,1)$ small enough
to guarantee the finiteness of the expectation; this is particularly important when the application of the Cauchy--Schwarz
inequality gives rise to the exponent $2s$.
The final bound on the Green functions can be extended to larger values
of $s$ with the help of the "one-for-all" principle (cf. Lemma \ref{lem:one.for.all}). In any case, the exponentially small
bounds on the fractional moments with any given $s>0$ imply exponential strong dynamical (and spectral) localization.

\vskip3mm
\noindent
\textbf{Step 1. Distant pairs of split configurations: no scale induction.}
If at least one of the configurations $\Bx$, $\By$ has a large diameter, the required decay bound follows -- without any scale induction --
from Lemma \ref{lem:exp.dH.diamx.diamy},  which takes now a simpler form, with $\diam \Pi\Bx \ge \diam \Pi\By$:
for some $A,m\in(0,+\infty)$
\be\label{eq:exp.dH.diamx.diamy}
\bigesm{ |\BG(\Bx,\By)|^s } \le A \exp\Big( - m\min \big[ \rdS(\Bx,\By), \; \diam \Pi \Bx \big]\Big).
\ee
Specifically,  if $\diam \Pi\Bx > L_k/2$, \eqref{eq:exp.dH.diamx.diamy} and \eqref{eq:dH.Lk.L.k+1} imply that
\be\label{eq:proof.main.Step1}
\bigesm{ |\BG(\Bx,\By)|^s } \le A \eu^{ - \frac{m}{2} L_k}.
\ee
Hence we can focus  in the rest of the proof on the pairs of configurations of diameter $\le L_k/2$.

\vskip1mm
\noindent
\textbf{Step 2. Decoupling inequality.} We start the analysis of the case where $\diam\, \Pi\Bx$, $\diam\,\Pi \By \le L_k/2$,
by establishing a decoupling inequaluty, essentially of the same nature as in the conventional, single-particle AM/FMM
approach (cf. \cite{AM93,ASFH01}). However, the decoupling achieved here is not as ``total'' as usual (i.e., for $N=1$), so one
looses the valuable mono-scale structure of the AM method and has to resort to a multi-scale procedure (see \textbf{Step 4}).

It is convenient to introduce the point $\Bzero := (0, 0)\in (\DZ^d)^2$.
Since $\Pi \Bx \ni 0$ and $\diam \Pi \Bx \le L_k/2$, we have $\Bx \in\bball_{L_k/2}(\Bzero)$; the latter cube is depicted as the
dashed square on Fig. 1.

At least one coordinate of  $\By$ equals $y$;
in the case $d=1$, $N=2$, possible positions of $\By$ are indicated on Fig.~1 by black dots, outside the
strip where $\diam(\cdot) \le L_k/2$, and by gray dots, inside the strip. In our argument, we are only concerned with the latter case
(gray dots on Fig. 1). Since we only know that $|y-0|>L_k$, the distance from $\By$ to $\bball_{L_k}(\Bzero)$ is not necessarily
large (it can be $=1$). Still, $\By$ is separated from $\Bx\in\bball_{L_k/2}(\Bzero)$ by the belt of width  $\ge L_k/2$,
and it is this belt that will provide the desired decay bound for the EFC.

By the GRE applied to the cube $\bball_{L_k}(\Bzero)$, with $\Bx$ inside and $\By$ outside it,
\be\label{eq:EFC.bound.GRE}
\bal
\BG(\Bx,\By) &= \sum_{(\Bw,\Bw')\in\pt \bball_{L_k}(\Bzero)} \BG_{\bball_{L_k}(\Bzero)}(\Bx,\Bw) \, \BDelta(\Bw,\Bw')\, \BG(\Bw',\By)
\\
& = \sum_{(\Bw,\Bw')\in\pt \bball_{L_k}(\Bzero)} \BG_{\bball_{L_k}(\Bzero)}(\Bx,\Bw) \, \BG(\Bw',\By),
\eal
\ee
yielding
$$
\hesm{ |\BG(\Bx,\By)|^s } \le \big| \pt \bball_{L_k}(\Bzero) \big|\,
   \max_{(\Bw,\Bw')\in\pt \bball_{L_k}(\Bzero) } \hesm{ |\BG_{\bball_{L_k}(\Bzero)}(\Bx,\Bw)|^s \, |\BG(\Bw',\By)|^s } .
$$

Here $\Pi\Bw' = \{w'_1, w'_2\}$ with $|0-w'_i|= L_k +1$ for at least one value
$i\in\{1,2\}$; we fix such $i$ and set $u_1=w'_i$, $u_2=y$.
Then $u_1, u_2 \not\in \ball_{L_k}(0)$.

Fix $(\Bw,\Bw')\in\pt \bball_{L_k}(\Bzero)$.
Introduce the sigma-algebra
$\fF_{\Lam\setminus \{u_1,u_2\}}$ generated by $\{V(z;\cdot), z\in\Lam \setminus \{u_1,u_2\}\}$ and
note that $\BG_{\bball_{L_k}(\Bzero)}(\Bx,\Bw)$ is $\fF_{\Lam\setminus \{u_1,u_2\}}$-measurable,
since $u_1, u_2 \not\in \ball_{L_k}(0)$, hence
$$
\bal
\hesm{ |\BG(\Bx,\By)|^s } &\le
\hesm{ |\BG_{\bball_{L_k/2}(\Bzero)}(\Bx,\Bw)|^s \; \hesm{ |\BG(\Bw',\By)|^s \cond \fF_{\Lam\setminus \{u_1,u_2\}} } }
\eal
$$
The conditional expectation in the above RHS is uniformly bounded, by virtue of
Lemma \ref{lem:apriori.EFC.bound} (cf. Eqn.~\eqref{eq:thm.apriori.EFC.bound}):
\be\label{eq:thm.apriori.EFC.bound.crossref}
\hesm{ |\BG(\Bw',\By)|^s \cond \fF_{\Lam\setminus \{u_1,u_2\}} } \le C_s |g|^{-s}  ,
\ee
thus
$$
\hesm{ |\BG(\Bx,\By)|^s } \le C_s |g|^{-s}  \; \hesm{ |\BG_{\bball_{L_k/2}(\Bzero)}(\Bx,\Bw)|^s  } .
$$

Assessing the above expectation is the most difficult task, and this hard work will be entrusted to the
scale induction (\emph{we carefully avoid using the words which would infringe the MSA's historical trademark}).

\vskip1mm
\noindent
\textbf{Step 3. Scaling step. I. Split configurations $\Bw$.}
We have  $\Pi \Bx \ni 0$, $\diam \Bx \le L_k/2$, so
the configuration $\Bx$ is "clustered" -- in the terminology of Ref.~\cite{AW09a}.

Consider first the case where $\Bw$ is ``split'':
$\diam \,\Pi\Bw > L_k/2$. Further, $|\Bw - \Bzero|=L_k$, $\Bzero$ is invariant w.r.t. the symmetry $(z_1,z_2)\mapsto (z_2,z_1)$,
so $\rdS(\Bw, \Bzero) = L_k$, while $\Bx\in\bball_{L_k/2}(\Bzero)$, thus we also have
$$
\rd_S(\Bx, \Bw)  \ge  L_k - \half L_k  =  \half L_k,
$$
and this situation is covered by Lemma \ref{lem:exp.dH.diamx.diamy}: with $\min[\rdS(\Bx,\Bw), \diam \Bx] \ge L_k/2$,
we have
$$
\hesm{|\BG_{\bball_{L_k}(\Bzero)}(\Bx,\Bw)|^{s} } \le \Const\, |g|^{-s}  \eu^{ - m \frac{L_k}{2}} .
$$

\vskip1mm
\noindent
\textbf{Step 4. Scaling step. II. Configurations $\Bw$ of restricted diameter.}
To single out the kind of EF correlators we will be working with, introduce the following
notation:
\be\label{eq:def.Ups}
\Ups(L) :=  \sup_{|I|\ge 1} \sum_{\Bx\in \bball_{L/2}(\Bzero)}
\sum_{\substack{\Bw\in\pt^- \bball_{L}(\Bzero) \\\diam \Bw \le L/2 }}
\hesmI{ |\BG_\Lam(\Bx,\Bw)|^s },
\ee
In the scaling procedure described below, it will be used with $L = L_{k+1}$.
We  also need a slightly
modified\footnote{Observe that, in the definition of $\Ups(L)$ with $L=L_{k+1}$, the diameter $L_k/2$
figuring in \eqref{eq:def.tUps} would have to be replaced by a larger one: $L_{k+1}/2$.}
quantity, defined for $L = L_{k+1}$, $k\ge 0$:
\be\label{eq:def.tUps}
\tUps(L_{k+1}) :=  \sup_{|I|\ge 1} \sum_{\Bx\in \bball_{ L_k/2 }(\Bzero)}
\sum_{\substack{\Bw\in\pt^- \bball_{L_{k+1}}(\Bzero) \\\diam \Bw \le L_k/2 }}
\hesmI{ |\BG_\Lam(\Bx,\Bw)|^s } .
\ee
$\Ups(L_k), \Ups(L_{k+1})$ are required to carry out the scale induction, but $\tUps(L_{k+1})$
is simpler to assess while performing the induction step.

We are going to show first that the quantities $\Ups(L_k)$, $k\ge 0$, satisfy the recursion
\be\label{eq:quad.recursion.claim}
\Ups\big(L_{k+1} \big) \le \frac{a}{|g|^s}\Ups^2\big(L_k \big) + A L_{k+1}^{2q} \eu^{- 2 \nu L_k},
\ee
and then infer from \eqref{eq:quad.recursion.claim} (cf. Lemma \ref{lem:quad.recursion}) that
$\Ups(L_k)$ decay exponentially.

\vskip1mm
\noindent
\textbf{(4.i)}
It is convenient to approximate $\Ups(L_{k+1})$ by $\tUps(L_{k+1})$.
Let us show that, for some $C$,
\be\label{eq:approx.Ups.tUps}
\bal
0 \le \Ups(L_{k+1}) - \tUps(L_{k+1}) &\le  C L_{k+1}^{2Nd} \,  \eu^{ - m \frac{L_k}{2} }.
\eal
\ee
The LHS inequality is obvious, since all the terms from $\tUps(L_{k+1})$ are present in $\Ups(L_{k+1})$.
Consider any term
$\hesm{ |\BG_\Lam(\Bx,\Bw)|^s }$
figuring in the sum for $\Ups(L_{k+1})$
but absent in $\tUps(L_{k+1})$ (cf. \eqref{eq:def.Ups}--\eqref{eq:def.tUps}). The exclusion from $\tUps(L_{k+1})$
implies the LHS inequality in
$$
 \half L_k < \diam \Pi\Bw \le  \half L_{k+1} ,
$$
while the RHS inequality is due to the constraint figuring in the definition of $\Ups(L_{k+1})$.
Further, $\Bw\in\pt^- \bball_{L_{k+1}}(\Bzero)$ implies that $\Pi \Bw \ni w$ with $|w - 0| = L_{k+1}$.
Since $\diam \Bx\le L_{k+1}/2$,  it follows
that
$$
\bal
\rdS(\Bx, \Bw) &= \min_{\pi\in\fS_N} \max_i | x_{\pi(i)} - w_i|  \ge \min_{j } | x_{j} - w |
\\
& \ge |0 - w| - | x_{j} - 0 | \ge L_{k+1} - \diam \Bx \ge  L_{k+1}  - \half L_{k+1}  > \half L_{k}.
\eal
$$
Now the lower bound $\diam \Bw > L_k/2$ allows us to apply Lemma \ref{lem:exp.dH.diamx.diamy} on $R$-distant configurations
at least one of which is $R$-split (here we have $R = L_k/2$):
$$
\bal
\hesm{ |\BG_\Lam(\Bx,\Bw)|^s } \le A \, \eu^{ - m \frac{L_k}{2} } ,
\eal
$$
so it remains only to assess the number of relevant terms, i.e., pairs $(\Bx,\Bw)$ figuring in \eqref{eq:def.Ups}.
We have $\Bx\in\bball_{L_{k+1}/2}(\Bzero)$ and $\Bw\in\pt^- \bball_{L_{k+1}}(\Bzero)$ with $\diam \Bw \le L_k/2$,
thus $\Bx, \Bw\in \bball_{3L_{k+1}/2}(\Bzero)$.
We conclude that the number of terms
which constitute the difference
$\Ups(L_{k+1}) - \tUps(L_{k+1})$ is bounded by
$CL_{k+1}^{2Nd}$. This completes the proof of \eqref{eq:approx.Ups.tUps}.

\vskip1mm
\noindent
\textbf{(4.ii)} Now we rescale the correlators $\Ups$, using in the process the reduced correlators $\tUps$.
This will be done with the help of a two-fold application of the FGRI.

Given the configurations $\Bx$ with $\diam \Bx\le L_k/2$ and $\Bw$ with $\diam \Bw \le L_k/2$,
we can assume w.l.o.g. that $\Pi\Bx \ni 0$ (otherwise we perform a diagonal shift
$(a,b) \mapsto (a+c, b+c)$
moving one of the particles in
$\Bx$ to $0\in\DZ^1$.  Let $\Pi\Bw \ni w$, $\hBw := (w, w)$.
By a two-fold application of the FGRI, setting for brevity $\bball' = \bball_{L_k}(\Bzero)$,
$\bball'' = \bball_{L_k}(\hBw)$ (observe that $L_k \approx L_{k+1}/2$),
\be\label{eq:three.GFs}
\bal
&\hesm{ |\BG(\Bx,\Bw)|^s}
\le \sum_{\substack{\lr{\Bu,\Bu'}\in\pt \bball' \\ \lr{\Bv,\Bv'}\in\pt \bball'' }}
  \hesm{ |\BG_{\bball' }(\Bx,\Bu)|^s \, |\BG(\Bu',\Bv')|^s   |\BG_{\bball'' }(\Bv,\Bw)|^s } .
\eal
\ee
Now we assess the middle factor in the RHS expectation.
Here $|\Bu' - \Bzero|=L_k+1$, so $\Pi\Bu'\ni u'$ with $|u' - 0|=L_k+1$,
and $u'\not\in\ball_{L_k}(0)$. Similarly,
$\Pi\Bv'\ni v'$ with $v'\not\in\ball_{L_k}(w)$.
(Cf. Fig. 2.)
Therefore both $\BG_{\bball' }(\Bx,\Bu)$ and $\BG_{\bball'' }(\Bv,\Bw)$
are measurable with respect to the sigma-algebra $\fF_{\ne u',v'}$ generated by all $V(z;\cdot)$ with $z\ne u', v'$.
At the same time, by Lemma \ref{lem:apriori.EFC.bound},
\be\label{eq:bound.cond.expect}
\hesm{|\BG(\Bu',\Bv')|^s \cond \fF_{\ne u', v'} } \le C |g|^{-s} < +\infty,
\ee
thus
$$
\hesm{ |\BG(\Bx,\Bw)|^s} \le \frac{C }{g^s} |\pt \bball'| |\pt \bball''|
\sum_{\substack{\lr{\Bu,\Bu'}\in\pt \bball' \\ \lr{\Bv,\Bv'}\in\pt \bball'' } }
    \hesm{|\BG_{\bball' }(\Bx,\Bu)|^s } \, \hesm{ |\BG_{\bball'' }(\Bv,\Bw)|^s}.
$$
\begin{center}
\begin{figure}
\begingroup
\centering

\begin{tikzpicture}
\begin{scope}[scale=0.59]
\clip (-4.9,-7.5) rectangle ++(14.5,17.5);

\draw[color=white] (-9, 0) circle (0.1);

\begin{scope}
\clip (-4.5,-4.0) rectangle ++(12.4,11.9);
\draw[color=gray!50!white!40, line width = 47] (-7.5,-7.5) -- (7.9,7.9);
\draw[color=lightgray, line width = 17] (-7.5,-7.5) -- (7.9, 7.9);
\end{scope}


\foreach \i in {   -4, -3.5, -3, -2.5, -2, -1.5, -1, -0.5, 0, 0.5, 1, 1.5, 2, 2.5, 3, 3.5, 4, 4.5, 5, 5.5, 6, 6.5, 7, 7.5  }
\draw[color=gray, line width = 0.5] (\i, -4.0) -- (\i, 7.9);

\foreach \i in {    -3.5, -3, -2.5, -2, -1.5, -1, -0.5, 0, 0.5, 1, 1.5, 2, 2.5, 3, 3.5, 4, 4.5, 5, 5.5, 6, 6.5, 7, 7.5  }
\draw[color=gray, line width = 0.5] (-4.4, \i) -- (7.9, \i);

\draw[color=black,, line width = 0.7, pattern=north west lines, pattern color=gray] (-1, -1) rectangle ++(2.0, 2.0);

\draw[color=black,, line width = 0.7] (-2, -2) rectangle ++(4.0, 4.0);

\draw[color=black,, line width = 0.7, pattern=north west lines, pattern color=gray] (4, 4) rectangle ++(2.0, 2.0);
\draw[color=black,, line width = 0.7] (3, 3) rectangle ++(4.0, 4.0);

\foreach \i in {-7, -6.5, -6, -5.5, -5, -4, -3.5, -3, -2.5, -2, -1.5, -1, -0.5, 0, 0.5, 1, 1.5, 2, 2.5, 3, 3.5, 4, 4.5, 5, 5.5, 6, 6.5, 7, 7.5 }
\fill (\i, \i) circle  (0.045);

\draw[color=black, line width = 1.2] (-7.5,0) -- (7.9, 0);
\draw[color=black, line width = 1.2] (0,  -4.0) -- (0, 7.9);

\fill[color=blue] (0.0, -1.0) circle  (0.15);

\fill[color=blue] (5.0, 4.0) circle  (0.15);

\fill[color=black] (5.0, 5.0) circle  (0.15);

\fill[color=black] (0.0, 0.0) circle  (0.15);

\fill[color=green!70!blue] (2.0, -1) circle  (0.10);
\fill[color=gray] (2.0, 0.5) circle  (0.10);

\draw[color=gray, line width =1] (2.5, 0.5) circle  (0.10);

\draw[color=green!70!blue, line width = 1] (2.5, -1) circle  (0.10);

\node at (0.75, -2.7) (Bx) {$\Bx$};
\draw[->,bend right=30, line width = 0.7] (Bx) to (0.2, -1.05);

\node at (1.7, -2.7) (Bu) {$\Bu$};
\draw[->,bend left=30, line width = 0.7] (Bu) to (1.9, -1.1);
\draw[->,bend left=30, line width = 0.7] (Bu) to (1.85, 0.45);

\node at (2.9, -2.7) (Bup) {$\Bu'$};
\draw[->,bend right=30, line width = 0.7] (Bup) to (2.55, -1.15);
\draw[->,bend right=30, line width = 0.7] (Bup) to (2.6, 0.4);

\node at (6.5, 2.7) (Bw) {$\Bw$};
\draw[->,bend left=20, line width = 0.7] (Bw.west) to (5.1, 3.8);

\node at (1.5, 5.7) (Bv) {$\Bv$};
\draw[->,bend left=30, line width = 0.7] (Bv) to (2.89, 5.15);

\fill[color=gray, line width =1] (3.0, 5.0) circle  (0.10);
\node at (1.0, 4.0) (Bvp) {$\Bv'$};
\draw[->,bend right=30, line width = 0.7] (Bvp) to (2.43, 4.85);

\draw[color=gray, line width =1] (2.5, 5.0) circle  (0.10);

\node at (1.1, 6.7) (hBw) {$\hBw$};
\draw[->,bend left=30, line width = 0.7] (hBw) to (4.85, 5.15);

\node at (-2.7, 1.5) (Bzero) {$\Bzero$};
\draw[->,bend left=30, line width = 0.7] (Bzero) to (-0.15, 0.15);

\draw[color=black, line width = 2] (-8.5,-5.5) -- (8.5,-5.5);
\draw[color=gray, line width = 3] (-2.0,-5.3) -- (2.0,-5.3);
\draw[color=gray, line width = 3] (3.0,-5.7) -- (7.0,-5.7);

\draw[color=green!70!blue, line width = 1] (2.5, -5.3) circle  (0.10);
\draw[color=gray, line width = 1] (2.5, -5.7) circle  (0.10);

\node at (-3.0,-4.7) (Lam0) {$\ball_{L_k}(0)$};
\draw[->, line width=1, bend left=30] (Lam0.east) to (-0.5,-5.1);

\node at (7.0,-7.0) (Lamw) {$\ball_{L_k}(w)$};
\draw[->, line width=1, bend left=30] (Lamw.west) to (4.5,-5.9);

\node at (5.5,-4.7) (up) {$u'$};
\draw[->, line width=1, bend right=30] (up) to (2.7,-5.2);

\node at (0.9,-7.0) (vp) {$v'$};
\draw[->, line width=1, bend right=30] (vp) to (2.4,-5.9);

\node at (9.3,-5.5) (DZ) {$\DZ^d$};

\end{scope}
\end{tikzpicture}

\endgroup

\caption{\emph{Example for the Step (4ii). The blue dots represent the configurations $\Bx$ and $\Bw$
figuring in the reduced correlators $\tUps$ (cf. \eqref{eq:def.tUps}). The points $\Bu, \Bu'$ and $\Bv,\Bv'$ are used in the FGRI.
The green dots $\Bu,\Bu'$ are the "split" configurations at the boundary, and the gray ones are
"clustered". Each of the configurations $\Bu'$ and $\Bv'$ contains at least one point
($u'$ and $v'$, respectively) outside the area $\ball_{L_k}(0)\cup \ball_{L_k}(w)\subset\DZ^d$.
This allows us to eliminate the Green function $\BG(\Bu',\Bv';E)$ in the RHS of Eqn.~\eqref{eq:three.GFs}
with the help of Lemma \ref{lem:apriori.EFC.bound},
thus decoupling the remaining factors. The dashed areas are the cubes of radius $L_k/2$ in $(\DZ^d)^2$.}}
\end{figure}
\end{center}
By the Hermitian symmetry of the Green functions,
both of the fractional  moments in the RHS can be assessed in the same way: by the appropriate diagonal shift,
leaving the expectation invariant,
one of the particles  in $\Bw$ can be moved to $0$, making the expectations
$\hesm{|\BG_{\bball'}(\Bx,\Bu)|^s}$ and $\hesm{|\BG_{\bball''}(\Bw,\Bu)|^s}$ similar. Thus it suffices to assess one of them, e.g.,
$\hesm{|\BG_{\bball'}(\Bx,\Bu)|^s}$.
This is done in two steps:
\begin{enumerate}[(a)]
  \item For $\Bu$ with $\diam\Bu \le L_k/2$ (see the green dots on Fig.2) we use the scale induction:
$$
\sum_{\substack{ \lr{\Bu,\Bu'}\in\pt \BLam_{L_k}(\Bzero)\\ \diam\Bu \le L_k/2}} \hesm{|\BG_{\bball'}(\Bx,\Bu)|^s} \le \Ups(L_k).
$$

  \item If $\diam\Bu > L_k/2$ (see the gray dots on Fig.2), then we use again Lemma \ref{lem:exp.dH.diamx.diamy}:
  with $N=2$,
$$
\hesm{|\BG_{\bball'}(\Bx,\Bu)|^s} \le A  \eu^{ - m  \frac{L_k}{2(N-1)} } = A  \eu^{ - \frac{m}{2} L_k }.
$$
\end{enumerate}

Collecting all possible vertices $\Bu$, falling into one of the above two categories (a) and (b),
and upper-bounding the number of vertices from each category by $|\pt^{-}\BLam_{}(\Bx)|$, we conclude that
$$
\sum_\Bu \hesm{|\BG_W(\Bx,\Bu)|^s} \le C |g|^{-s} \left( \Ups(L_k) +  C' L_k^{q} \eu^{ - m L_k }\right),
\;\ q = 2Nd-2 = 2.
$$
Similarly,
$$
\sum_\Bv \hesm{|\BG_V(\Bv,\Bw)|^s} \le C |g|^{-s} \big( \Ups(L_k) +  C'  L_k^{q} \eu^{ - \frac{m}{2} L_k } \big).
$$
Therefore,
$$
\tUps(L_{k+1}) \le C |g|^{-s} \left( \Ups(L_k) +  C' L_k^{q} \eu^{ - \frac{m}{2} L_k }\right)^2.
$$
Finally, on account of the approximation formula \eqref{eq:approx.Ups.tUps}, we obtain
$$
\Ups(L_{k+1}) \le C |g|^{-s} \left( \Ups(L_k) +  C' \eu^{ - \frac{m}{2} L_k }\right)^2 +
 A L_{k+1}^{p} \eu^{ - \frac{m}{2} L_k} .
$$
and with $\hUps_k := \Ups(L_k) +  C' \eu^{ - \frac{m}{2} L_k }$, $A' = 4 \max(A, C')$, $M_s := 2C |g|^{-s}$,
we come to the recursive inequality
\be\label{rec.hUps}
\hUps_{k+1} \le \half M_s \hUps^2_k +  \half A' L_{k+1}^{p} \eu^{ - \frac{m}{2} L_k} .
\ee

\noindent
An elementary
calculation\footnote{A reader familiar with the work \cite{GK01} by Germinet and Klein can recognize some elements
of the proof of \cite{GK01}*{Theorem 5.1} (cf. in particular \cite{GK01}*{Eqn. (5.30)}
in the argument used in Appendix \ref{app:quadratic.dyn}.}
(cf. Lemma \ref{lem:quad.recursion}) shows that the recursion \eqref{rec.hUps} implies
$$
\forall\, k\ge 0 \quad \Ups(L_k) \le \Const(|g|, s, m) \,  \eu^{ - \tmu L_k },
$$
with $\tmu >0$. For brevity, we omit here the explicit expression for the lower bound on $\tmu$;
see the details in Appendix \ref{app:quadratic.dyn}.

\vskip1mm
\noindent
\textbf{Step 5. Conclusion.}

Given two configurations $\Bx,\By\in\Lam^2$ with $\rdS(\Bx,\By) =D$, there exist $\tmu>0$ such that

$$
\hesm{ |\BG(\Bx,\By;E)|^s } \le
\left\{
  \begin{array}{ll}
    \Const\, \eu^{- \frac{m}{2} D}, & \hbox{ if $ \diam (\Bx) \vee \diam (\By) > D/2$;} \\
    \Const\, \eu^{- \tmu D}, & \hbox{ if $\diam (\Bx), \diam (\By) \le D/2$.}
  \end{array}
\right.
$$

Thus the assertion of Theorem \ref{thm:Main.finite} is proved.
\qed

\vskip3mm

Recall that the proof makes use of two important estimates, Lemma \ref{lem:apriori.EFC.bound}
on the finiteness of fractional moments, and Lemma \ref{lem:exp.dH.diamx.diamy}
addressing the particle transfer processes to/from split configurations. Their proofs
do not use the scale induction, and this is one of the reasons we prove them separately.

\section{Proof of Theorem \ref{thm:Main.infinite}}

A direct inspection of the proof of Theorem \ref{thm:Main.finite} shows that the finiteness of the range
of interaction is used only in the proof of Lemma \ref{lem:exp.dH.diamx.diamy}, providing an exponential
upper bound on the fractional moment of the GFs $\BG(\Bx,\By;E)$ where at least one of the configurations
$\Bx$, $\By$ is "split", and the two are "distant". Further, Lemma \ref{lem:exp.dH.diamx.diamy}
is proved without scale induction, as a separate statement logically independent of the rest
of the proof of Theorem \ref{thm:Main.finite}. Therefore, it suffices to prove an analog of
Lemma \ref{lem:exp.dH.diamx.diamy} for the exponentially decaying interactions of infinite range,
and such a statement (Lemma \ref{lem:exp.dH.diamx.diamy.2}) is proved in Section \ref{sec:split.infinite}.

Replacing the statement of Lemma \ref{lem:exp.dH.diamx.diamy} with that of Lemma \ref{lem:exp.dH.diamx.diamy.2} in the arguments
given in Section \ref{sec:proof.Main.finite}, the assertion of Theorem \ref{thm:Main.infinite} follows.
\qed

\section{Finiteness of the fractional moments}
\label{sec:finiteness.FM}

A finite a priori bound on the fractional moments of the resolvents
is one of the inescapable ingredients of the FMM, going back to the pioneering paper \cite{AM93}.
Its adaptation to the multi-particle models (cf. \cite{AW09a}) is, however, more involved than in the $1$-particle theory.
Both Ref.~\cite{AW09a} and a more recent work \cite{FW14},
as well as Ref.~\cite{AENSS06} dedicated to the FMM for differential random operators,
refer to some general results on maximally dissipative
operators and related topics;
cf., e.g., \cite{As67}, \cite{AENSS06}, \cite{BirE67}, \cite{DeB62}, \cite{CHN01}, \cite{HSim02}, \cite{Nab91},
and the monograph \cite{Stein70}.
This list (certainly incomplete)  constitutes a highly recommended introductory reading (300+ pages).
In the present manuscript, however, we restrict our analysis to the lattice models,
and actually work only with finite-dimensional operators, so it is possible to give an upshot of the required theory
"in a nutshell", with a bare minimum of elementary technical tools. In fact, it all boils down to a straightforward application
of the Boole formula. The reader can see, e.g. in \cite{AENSS06}, that a more general situation requires
far-going generalizations of this simple identity.

\ble\label{lem:apriori.EFC.bound}
For any $s\in(0,1)$ there exists $C_s<\infty$ such that for any finite connected subset $\Lam\subset\DZ^d$, any
two sites $u_1, u_2\in\DZ^d$, with $n := \card\, \{u_1, u_2\} \in\{1,2\}$, and any pair of configurations $\Bx, \By\in \Lam^2$
with $\Pi\Bx \ni u_1$, $\Pi\By \ni u_2$, the following bound holds: some $C=C(s, F_V, n )<\infty $,
\be\label{eq:thm.apriori.EFC.bound}
\hesm{ |\BG(\Bx,\By;E)|^s \cond \fF_{\ne u_1,u_2} } \le  C |g|^{-s },
\ee
where $\fF_{\ne u_1,u_2}$ is the sigma-algebra generated by
$\{V(u;\cdot), \, u\in\DZ^d\setminus\{u_1, u_2\}\}$.
\ele

\proof
\textbf{I.}
We consider first the case where $u_1\ne u_2$.

As we shall see, the relevant representation of the random operator at hand is $g\BV(\om) + \BA$, with the nonrandom
component $\BA = \BH_0 + \BU$, and we work with the resolvent $\BG_g(E)=(g\BV(\om) + \BA - E)^{-1}$. The random field
$V$ is assumed bounded, $\|V(x;\cdot)\|_\infty<+\infty$, and it suffices to assume that $\|V(x;\cdot)\|_\infty\le 1$,
for larger values are simply obtained by taking $|g|$ larger. In fact, even the particular model where $V\sim \Unif([0,1])$
is of great interest, and it is one of the most popular models of disorder in physics. Then we can extract the factor $g$
and note that, with $\BB_{g} := g^{-1}\BA$, $\lam= g^{-1} E$,
$$
\hesm{ |\BG_g(\Bx,\By;E)|^s \cond \fF_{\ne u_1,u_2} }
= |g|^{-s} \, \hesm{ (\one_\By, \big( \BV(\om) + \BB_{g} - \lam \big)^{-1} \one_\Bx )} .
$$
In the rest of the proof, we work with the resolvent of the operator $\BV(\om) + \BB_{g}$, at a rescaled
energy $\lam$ (which is fixed in the proof, anyway).

\noindent
\textbf{$\bullet$ Reduced probability space.} Now the r.v. $V(x;\om)$ vary inside $I=[0,1]$ and admit a
bounded probability density $p_V$,
$\| p_V \|_\infty =\bp<\infty$. The conditional distribution
of $V$ given $\fF_{\ne u_1,u_2}$ gives rise to the reduced probability space
$(A, \tDP)$, where $A = I^2$, $\tDP$ is absolutely continuous
with respect to the Lebesgue measure $\mes_I\otimes\mes_I$ on  $A$, with density
$(v_1, v_2)\mapsto \bp(v_1,v_2) = p_V(v_1) \, p_V(v_2) \le \bp^2$. Then for
any non-negative random variable $\tzeta$ on $(\Om,\DP)$ and any $s\in(0,1)$, the conditional expectation of $\tzeta^s$
given $\fF_{\ne u_1,u_2}$ has the form (below we allow the expectation to be $+\infty$)
\be\label{eq:FM.cond.ne}
\bal
0 \le \hesm{ \tzeta^s(\om) \cond \fF_{\ne u_1,u_2} } &= \int_I dv_1 \int_I dv_2 \; p_V(v_1) \, p_V(v_2)\, \zeta^s(v_1, v_2; \cdot)
\\
& \le \bp^2  \int_{A} dv_1 \, dv_2\, \zeta^s(v_1, v_2; \bullet),
\eal
\ee
where $\zeta(v_1, v_2; \bullet)$ is obtained from $\tzeta(\om)$ by identifying $v_j \equiv V(u_j;\om)$, with the remaining
degrees of freedom fixed by conditioning (they are symbolically represented by $\bullet$). Now $\zeta$ can be considered as
a random variable on the square $I^2$ with the normalized Lebesgue measure. For the rest of the argument, $\pr{}$ and
$\hesm{}$ refer to this new probability space. Let $F_\zeta(t) = \pr{\zeta \le t}$, then
\be\label{eq:integral.zeta.s}
\hesm{ \zeta^s} = \int_{A} dv_1 \, dv_2\,  \zeta^s(v_1, v_2; \cdot)
= \int_0^\infty t^s dF_\zeta(t) = s\int_0^\infty t^{s-1} (1 - F_\zeta(t))\, dt.
\ee
We shall return to \eqref{eq:integral.zeta.s}, once we obtain a suitable upper bound of the tail probability distribution
function (below $\mes$ is the Lebesgue measure on $A$)
$$
1 - F_\zeta(t) =  \mes\{ (v_1,v_2)\in A:\; \zeta(v_1,v_2) >t\}.
$$

\noindent
\textbf{$\bullet$ The Birman--Schwinger relation.}
Introduce the sets
$$
\BS_j = \Big(\big( \{u_j\}\times \DZ\big) \cup \big(\DZ\times\{u_j\} \Big)\cap \Lam, \;\; j=1,2,
$$
the (multiplication) operators
$$
0 \ne \BC = \one_{\BS_1} + \one_{\BS_2} \ge 0,   \;\;     \BD = \one_{\BS_1} - \one_{\BS_2},
$$
and the random variables
$$
\xi = \half( V(x_1;\om) + V(x_2;\om) ), \;\; \eta = \half( V(x_1;\om) - V(x_2;\om) ).
$$
Then
$$
\bal
V(u_1;\om) \one_{\BS_1} + V(u_2;\om) \one_{\BS_2} &= \xi \BC + \eta \BD
\eal
$$
and
$$
\bal
\BH(\om) &= \tBK(\om) + g V(u_1;\om) \one_{\BS_1} + g V(u_2;\om) \one_{\BS_2},
\\
& = \tBK + g \xi \BC + g\eta \BD = \BK + g \xi \BC,
\eal
$$
where $\tBK(\om)$ is $\fF_{\ne u_1,u_2}$-measurable and $\BK(\om) = \tBK(\om) + \eta(\om)\BD$.

The operator $\BC$ is non-negative and not identically zero, so we can use the Birman--Schwinger identity
for $\BK_E = \BK - E$:
$$
\BC^{1/2} (\BK_E  + g\xi \BC)^{-1} \BC^{1/2} = \left( \BC^{1/2} \BK_E^{-1} \BC^{1/2} + g\xi\one \right)^{-1} .
$$
The operator
$$
\BK_{E,\BC} := \BC^{1/2}(\BK_E  + g \xi \BC)^{-1}\BC^{1/2}
$$
is considered acting in the subspace of $\cH$,
$$
 \cH_{\{u_1,u_2\}} = \left( \Ker \BC \right)^\perp =
 \spann \{ \one_{\Bw}:\, \Bw\in\Lam^2,\,  \Pi\Bw \cap \{u_1,u_2\}\ne \varnothing \},
$$
containing in particular $\one_{\Bx}$ and $\one_{\By}$.
Its relevance is explained by the fact that both $\one_{\Bx}$ and $\one_{\By}$ are
eigenvectors of $\BC$ with positive eigenvalues,
$$
\BC \one_\Bx = \one_{\BS_1} \one_\Bx + \one_{\BS_2} \one_\Bx = \big(\BN_{u_1}(\Bx) + \BN_{u_2}(\Bx) \big) \one_\Bx
= \alpha_\Bx \one_\Bx,
$$
with
$$
\BN_w(\Bu) := \card\{j\in\{1,2\}: \, u_j = w\}
$$
(the number of particles in $\Bu$ at the position $u$),
hence
\be\label{eq:B.alpha}
\bal
\BC^{1/2} \one_\Bx &= \alpha^{1/2}_\Bx \one_\Bx,
\\
\BC^{1/2} \one_\By &= \alpha^{1/2}_\By \one_\By,
\eal
\ee
with $1 \le \alpha_\Bx, \alpha_\By \le 2$. Therefore, with $\alpha := (\alpha_\Bx \alpha_\By)^{-1/2}\in[1,1/2]$,
$$
\bal
\BG(\Bx,\By;E) &=
(\one_\By, (\BK_E  + g \xi \BC)^{-1} \one_\Bx)
\\
&= \alpha\,  (\one_\By, \BC^{1/2}(\BK_E  + g \xi \BC)^{-1}\BC^{1/2} \one_\Bx)
\\
& = \alpha
\left( \one_\By, \left(\BC^{1/2} \BK_E^{-1} \BC^{1/2} + g \xi\one \right)^{-1} \one_\Bx \right)
\eal
$$

Since $\alpha \le 1$, one has an implication: for any $t>0$,
$$
|\BG(\Bx,\By;E)| > t \Longrightarrow
\left| \, \left( \one_\By, \left(\BK_{E,\BC} + g \xi\one \right)^{-1} \one_\Bx \right) \,\right| > t,
$$
where the RHS refers to the (finite-dimensional) space $\cH_{\{u_1,u_2\}} $.

\vskip1mm
\noindent
\textbf{$\bullet$ The tail tale and the Boole formula.}
Consider the linear change of variables
$\Phi:(v_1,v_2) \mapsto (\xi,\eta) = ((v_1+v_2)/2, (v_1-v_2)/2)$ with Jacobian $=2$.
Note that with $(v_1,v_2)\in A$, $\xi$ varies in $[0,1]$ and $\eta$ in $[-1/2,1/2]$.

Let $A'=\Phi(A)\subset [0,1] \times [-1/2,1/2]$,
then by the Fubini theorem,
\be\label{eq:int.R2}
\bal
\int_{\DR^2} dv_1\, dv_2\, \one_A \one_{M_t}
&= 2 \int_{-1/2}^{1/2} d\eta\, \int_{\DR} d\xi\, \one_{A'}(\xi,\eta) \one_{M_t}\circ \Phi^{-1}(\xi,\eta)
\\
& \le 2 \int_{-1/2}^{1/2} d\eta\, \int_{\DR} d\xi\, \one_{M_t}\circ \Phi^{-1}(\xi,\eta)
\\
& \le 2 \cdot  \sup_{\eta\in\DR} \; \mes\big( \cM_{t}(\eta) \big) ,
\eal
\ee
where
$$
\cM_{t}(\eta) = \left\{\xi\in\DR:\,
\left| \, \left( \one_\By, \left(\BK_{E,\BC,\eta} + g \xi\one \right)^{-1} \one_\Bx \right) \,\right| > t
\right\} .
$$
The function
$$
\cR: \xi \mapsto \left( \one_\By, \left(\BK_{E,\BC,\eta} + g\xi\one \right)^{-1} \one_\Bx \right)
\equiv g^{-1} \left( \one_\By, \left(g^{-1}\BK_{E,\BC,\eta} + \xi\one \right)^{-1} \one_\Bx \right)
$$
is rational, with real simple poles,
$$
\cR(\xi) = \sum_j \frac{g^{-1} c_j}{ \lam_j - \xi}, \;\; \sum_j |g^{-1}c_j| \le |g|^{-1} \;\text{ (by Bessel's inequality)},
$$
so we can again apply the Boole formula,
$$
\mes\{ \xi:\, |\cR(\xi)| > t\} = \frac{2\sum_i c_i}{t} \le \frac{2}{g t}.
$$
Therefore,
$$
1 = \mes A \ge \int_{\DR^2} dv_1\, dv_2\, \one_A \one_{M_t} \le 2  \cdot \frac{2}{ gt} = \frac{4}{g t},
$$
yielding for the tail probability distribution function
$$
1 - F_\zeta(t) \le  \min\big[ 1, 4 g^{-1} t^{-1}  \big] .
$$

\vskip1mm
\noindent
\textbf{$\bullet$ The fractional moment.}
Return to the fractional moment in \eqref{eq:integral.zeta.s}:
$$
\bal
\hesm{ \zeta^s } & =
s \int_0^\infty t^{s-1} (1 - F_\zeta(t))\, dt
\\
 &\le s \int_0^\infty t^{s-1} \, \min(1, 4 g^{-1} t^{-1}) \, dt
\\
& = s\int_0^{4/g} t^{s-1}\, dt + s\int_{4/g}^{+\infty} t^{s-2}\, dt
\\
& = \frac{4^s }{g^s (1-s)}  .
\eal
$$
Finally,
for the conditional fractional moment in \eqref{eq:thm.apriori.EFC.bound}, we obtain, as asserted,
$$
\hesm{ |\BG(\Bx,\By;E)|^s \cond \fF_{\ne u_1,u_2} } \le \bp^2  \hesm{ \zeta^s }  \le \frac{C}{ |g|^s(1-s)} .
$$

\vskip3mm
\noindent
\textbf{II.}
Now we turn to the simpler case where $u_1=u_2 = u$. The two-parameter operator families introduced above
are controlled now with a single parameter. Consider the set
$$
\BS =  \big(\{u\} \times \DZ) \cup (\DZ  \times \{u\}) \big) \cap \Lam,
$$
and let $\BC=\one_\BS$; then $\BH = g\xi \BC + \BK(\om)$, where $\BK$ is $\fF_{\ne u}$-measurable.

The operator $\BC$ is non-negative and nonzero, so we can use the Birman--Schwinger identity
for $\BK_E = \BK - E$:
$$
\BC^{1/2} (\BK_E  + g \xi \BC)^{-1} \BC^{1/2} = \left( \BC^{1/2} \BK_E^{-1} \BC^{1/2} - g \xi\one \right)^{-1} .
$$
The functions $\one_{\Bx}$, $\one_{\By}$ are eigenvectors of $\BC$ with positive eigenvalues, viz.
$$
\BC \one_\Bx = \one_{\BS} \one_\Bx = \BN_{u}(\Bx) \one_\Bx,
\BC \one_\By = \one_{\BS} \one_\By = \BN_{u}(\By) \one_\By,
$$
with $\BN_w(\Bu) := \card\{j\in\{1,2\}: \, u_j = w\}$,
hence
$$
\bal
\BC^{1/2} \one_\Bx &= \alpha^{1/2}_\Bx \one_\Bx, \;\;
\BC^{1/2} \one_\By &= \alpha^{1/2}_\By \one_\By,
\eal
$$
with $1 \le \alpha_\Bx, \alpha_\By \le 2$. Therefore, as in \eqref{eq:B.alpha}, we obtain
$$
\bal
\BG(\Bx,\By;E) &=
(\one_\By, (\BK_E  + g \xi \BC)^{-1} \one_\Bx)
 = \alpha
\left( \one_\By, \big(\BC^{1/2} \BK_E^{-1} \BC^{1/2} + g \xi\one \big)^{-1} \one_\Bx \right).
\eal
$$
The operator
$
\BK_{E,\BC} := \BC^{1/2}(\BK_E  + g \xi \BC)^{-1}\BC^{1/2}
$
acts  in the subspace of $\cH$,
$$
 \cH_{\{u\}} = \left( \Ker \BC \right)^\perp =
 \spann \{ \one_{\Bw}:\, \Bw\in\Lam^2,\,  \Pi\Bw \cap \{u\}\ne \varnothing \},
$$
containing $\one_{\Bx}$ and $\one_{\By}$.
Since $\alpha \ge 1$, for any $t>0$,
$$
|\BG(\Bx,\By;E)| > t \Longrightarrow
\left| \, \left( \one_\By, \left(\BK_{E,\BC} + g \xi\one \right)^{-1} \one_\Bx \right) \,\right| > t .
$$
Introduce the rational function
$\cR: \xi \mapsto \big( \one_\By, \left(\BK_{E,\BC,\eta} + g \xi\one \right)^{-1} \one_\Bx \big)$,
$$
\cR(\xi) = \sum_j \frac{g^{-1} c_j}{ \lam_j - \xi}, \;\; \sum_j |g^{-1}c_j| \le g^{-1} \;\text{ (by Bessel's inequality)},
$$
and the set
$ \cM_{t} = \left\{\xi\in I:\, \big|\cR(\xi) \big| > t \right\}$
with $\mes \cM_t \le 1$.
By the Boole inequality combined with the LHS inequality in \eqref{eq:int.R2},
$$
 \mes\{ \xi:\, |\cR(\xi)| > t\} \le \min\big[ 1, 2 g^{-1}t^{-1} \big].
$$
Hence
$$
\bal
\hesm{\zeta^s} & \le \int_0^\infty t^s dF_\zeta(t) = s\int_0^\infty t^{s-1} (1 -F_\zeta(t) ) \, dt
\\
& \le s\int_0^\infty t^{s-1} \min\big[ 1, 2 g^{-1}t^{-1} \big] \, dt
= \frac{ 2^s }{g^s(1-s)}
\eal
$$
so the original (conditional) fractional moment in \eqref{eq:thm.apriori.EFC.bound} is bounded:
$$
\hesm{ |\BG(\Bx,\By;E)|^s \cond \fF_{\ne u_1,u_2} } \le \frac{C }{ g^s(1 - s) }.
$$
\qed

\section{Tunneling from split configurations. Finite-range interaction}
\label{sec:split.finite}

\ble\label{lem:exp.dH.diamx.diamy}
Assume that the interaction potential has finite range $r_0$.
There exist some $A =A(d, r_0), m'>0$ such that if the $1$-particle system is $m$-localized with $m\ge 1$, then
\be\label{eq:exp.dH.diamx.diamy.again}
\bigesm{ |\BG(\Bx,\By)|^s } \le A \exp\Big( -m' \min \big[ \rdS(\Bx,\By), \; \diam(\Bx) \vee \diam(\By) \big] \Big).
\ee
\ele

\proof
The main argument benefits from the assumption $N=2$ which simplifies the combinatorial analysis, as well as notation,
since there is only one way to split a $2$-particle configuration into two distant "clusters" -- single-particle
sub-configurations, in this case. Here $\Pi \Bx =\{x_1, x_2\}$, $x_1, x_2\in \DZ^d$,
$\diam \Bx = |x_1 - x_2|$, and we assume that
$$
\rdS(\Bx, \By), |x_1 - x_2| \ge R >0.
$$
Thus either $\dist(x_i, \Pi\By) \ge R$ for some $i\in\{1,2\}$, or
$\dist(y_j, \Pi\Bx) \ge R$ for some $j\in\{1,2\}$. In either case,
\be\label{eq:max.dist.x.y}
|x_1 - y_1| \vee |x_2 - y_2| \ge \rdS(\Bx,\By) \ge R.
\ee
In the proof, we  treat $\BU$ as a perturbation of the Hamiltonian $\BHni$.
With $\BH = \BHni + \BU$, the FGRI gives
\be\label{eq:GRE.perturb.U}
|\BG(\Bx,\By)|^s \le |\BGni(\Bx,\By)|^s + |\big(\BGni \BU \BG\big)(\Bx,\By)|^s .
\ee

\noindent
\textbf{Step 1.}
Let us show that
\be\label{eq:EFC.non-int}
\forall\, \Bu,\Bv\in\DZ^2 \quad
\hesm{ |\BGni(\Bu,\Bv)|^{2s} } \le A \eu^{ - m\, \rdS(\Bu,\Bv)}.
\ee
This general bound will be first applied to the term $\BGni(\Bx,\By)$ in \eqref{eq:GRE.perturb.U}, and
later, at \textbf{Step 2}, to the expectations $\BGni(\Bx,\Bw)$ in \eqref{eq:esm.eps.R}.

By assumption on of the $1$-particle systems,
$$
Q_\Lam^{(1)}(x, y; \DR) \le A \eu^{- m\, |x-y|}.
$$
For any interval $I\subset \DR$ we have a deterministic relation between the $2$-particle
and $1$-particle EFC:
$$
\bal
\BQni(\Bu,\Bv; I) &= \sum_{\substack{\lam_i\in\Sigma_1,\, \mu_j\in\Sigma_2 \\ \lam_i+\mu_j\in I} }
| \ffi_i(x_1) \overline{\ffi}_i(x_1) \, \psi_j(x_2) \overline{\psi}_j(y_2)|
\\
& \le \sum_{\lam_i\in\Sigma_1}
| \ffi_i(x_1) \overline{\ffi}_i(y_1) |
\sum_{ \mu_j\in\Sigma_2  }
|  \psi_j(x_2) \overline{\psi}_j(y_2)|
\\
& = Q^{(1)}(x_1,y_1 ) \, Q^{(2)}(x_2,y_2).
\eal
$$
Thanks to the a priori deterministic upper bound on the EFC, $Q(u,v;E)\le 1$, we obtain,
on account of \eqref{eq:max.dist.x.y},
\begin{align}
\label{Qni.1}
\hesm{ (\BQni(\Bu,\Bv;E)\big)^s} &\le \min \big\{ \hesm{Q^{(1)}(x_1,y_1 )}, \; \hesm{Q^{(2)}(x_2,y_2 )} \big\}
\\
\label{Qni.2}
& \le A \exp\left( - m \, \max \big[ |x_1 - y_1|, \, |x_2 - y_2| \big] \right)
\\
\label{Qni.3}
& \le A \eu^{  - m \,\rdS(\Bu,\Bv) } .
\end{align}
We conclude this stage of analysis by applying Lemma \ref{lem:EFC.to.GF}:
$$
\bal
\hesm{ |\BGni(\Bu,\Bv)|^s} &\le \frac{C }{1-s} \hesm{ |\BQni(\Bu,\Bv)|^s}
\le \frac{C }{1-s} A \eu^{  - m \,\rdS(\Bu,\Bv) } .
\eal
$$

\vskip2mm
\noindent
\textbf{Step 2.}
Now consider the second, perturbative term in \eqref{eq:GRE.perturb.U}.
Since $\BU$ is
diagonal in the delta-basis, and the support of the interaction energy function,
$\Bw \mapsto \BU(\Bw)$, is contained in the strip $\DD_{r_0} := \{\Bz\in (\DZ^d)^2:\, \diam \Bz \le r_0\}$, we have
$$
\bal
\eps_R &= \eps_R(\om) :=
\| \BU\|^{-1} \, |\big(\BGni \BU \BG\big)(\Bx,\By)|^s
\\
& \le \| \BU\|^{-1} \,\sum_{\Bw\in\DD_{r_0}} |\BGni(\Bx,\Bw)|^s\,  |\BU(\Bw|^s)\, |\BG(\Bw,\By)|^s
\\
& \le   \sum_{\Bw\in \DD_{r_0}} |\BGni(\Bx,\Bw)|^s\, \, |\BG(\Bw,\By)|^s.
\eal
$$
By the Cauchy--Schwarz inequality,
\be\label{eq:esm.eps.R}
\bal
\hesm{ \eps_R } & \le \sum_{\Bw\in\DD_{r_0}} \left( \hesm{ |\BGni(\Bx,\Bw)|^{2s} } \right)^{1/2} \,
  \left( \hesm{ |\BG(\Bw,\By)|^{2s} } \right)^{1/2}.
\eal
\ee
The last fractional moment in the RHS of \eqref{eq:esm.eps.R} does not provide any significant contribution to the \emph{decay} bound,
for there is little we assumed about $\By$ (except that $\rdS(\Bx,\By) = R>L_k$). Indeed, we merely claim that it is "harmless",
making use of the results of Sect.~\ref{sec:finiteness.FM}: with $0 < 2s < 1$,
\be
\hesm{ |\BG(\Bw,\By)|^{2s} } \le \frac{\Const}{(1-2s)  |g|^{2s} }.
\ee
Therefore,
$$
\bal
\hesm{ \eps_R } & \le  \frac{\Const}{(1-2s)  |g|^{2s} } \sum_{\Bw\in\DD_{rr_0}} \left( \hesm{ |\BGni(\Bx,\Bw)|^{2s} } \right)^{1/2} .
\eal
$$
Applying again the general bound \eqref{eq:EFC.non-int} with $(\Bu,\Bv) = (\Bx,\Bw)$,
we can write
\be\label{eq:sum.Bw.supp.U}
\bal
\sum_{\Bw\in\supp \BU}
\hesm{ |\BGni(\Bx,\Bw)|^{2s} } &\le A  \sum_{\Bw\in\supp \BU} \eu^{ - m\, \rd(\Bx,\Bw)} .
\eal
\ee
For $\Bw\in\supp \BU$, the distance $\rd(\Bx,\Bw)$ is essentially the distance from
$\Bx$ to the diagonal $\DD_0 = \{(u,u), u\in\DZ^d\}$. To be more precise,
consider a point $\Bw\in\supp \BU$, thus with  $\diam \Bw \le r_0$, and let
$$
 \rdS(\Bx,\Bw)= \max\big[ |x_1 - w'|, |x_2 - w''| \big], \;\; \{w', w''\} = \Pi \Bw.
$$
Then by the triangle inequality,
$$
r_0 \ge \diam\Bw = |w' - w''| \ge |x_1 - x_2| - |x_1 - w'| - |x_2 - w''| ,
$$
hence
\be\label{eq:lower.bound.dist.Bx.Bw.finite}
\rdS(\Bx,\Bw) \ge \left \lceil\frac{\diam \Bx - r_0}{ 2 } \right\rceil \ge \frac{\diam \Bx - r_0}{ 2 } =:R_\Bx
\ge \half(R - r_0) .
\ee
Now we can apply crude upper bounds
\be
\card\{ \Bw\in(\DZ^d)^2: \,  \rdS(\Bx, \Bw) = r \} \le C r^{2d},
\ee
and
\be
\bal
 \sum_{\Bw\in\supp \BU} \eu^{ - m\, \rd(\Bx,\Bw)} &\le  \sum_{\Bw\in\supp \BU} \eu^{ - m\, \rd(\Bx,\Bw)}
\\
& \le  C \sum_{r \ge R_\Bx } r^{2d} \eu^{ - m\, r } \le
  C (R_\Bx)^{2d} \eu^{ - m\, R_\Bx }
\\
& \le C' R^{2d} \eu^{ - \frac{m}{2} R } .
\eal
\ee
Concluding, we obtain
\be\label{eq:esm.eps.R.conclude}
\hesm{ \eps_R } \le \Const\, |g|^{-s } \eu^{ - \frac{m}{2}\, R}.
\ee
Collecting \eqref{eq:GRE.perturb.U}, \eqref{eq:EFC.non-int} with $(\Bu,\Bv)=(\Bx,\By)$,
and  \eqref{eq:esm.eps.R.conclude}, the claim follows.
\qedhere


\section{Tunneling from split configurations. Infinite-range interaction}
\label{sec:split.infinite}

\ble\label{lem:exp.dH.diamx.diamy.2}
Suppose that
$$
\min \big[ \rdS(\Bx,\By), \; \diam(\Bx) \vee \diam(\By) \big] \ge R >0.
$$
Then for some $m'>0$, one has
\be\label{eq:exp.dH.diamx.diamy.2}
\bigesm{ |\BG(\Bx,\By)|^s } \le A \eu^{ -m' R }.
\ee
\ele

\proof
We assume that
$$
\rdS(\Bx, \By), |x_1 - x_2| \ge R >0.
$$
Thus either $\dist(x_i, \Pi\By) \ge R$ for some $i\in\{1,2\}$, or
$\dist(y_j, \Pi\Bx) \ge R$ for some $j\in\{1,2\}$. In either case,
\be\label{eq:max.dist.x.y.2}
|x_1 - y_1| \vee |x_2 - y_2| \ge \rdS(\Bx,\By) \ge R.
\ee

In the proof, we  treat $\BU$ as a perturbation of the Hamiltonian $\BHni$; more precisely, we assess the effect
of $\BU$ on the fractional moment of the Green function $\BG(\Bx,\By)$, with $\Bx$ far away from the support
of the interaction potential, and with $\Pi \By$  distant from $\Pi \Bx$. The geometric
analysis would be more involved for $N\ge 3$.

With $\BH = \BHni + \BU$, the GRI gives
\be\label{eq:GRE.perturb.U.2}
|\BG(\Bx,\By)|^s \le |\BGni(\Bx,\By)|^s + |\big(\BGni \BU \BG\big)(\Bx,\By)|^s .
\ee

\noindent
\textbf{Step 1.}
As shown in the proof of Lemma \ref{lem:exp.dH.diamx.diamy}, the Green functions of $\BHni(\om)$ satisfy
\be\label{eq:EFC.non-int.2}
\forall\, \Bu,\Bv\in(\DZ^d)^2 \quad
\hesm{ |\BGni(\Bu,\Bv)|^{2s} } \le A \eu^{ - m\, \rdS(\Bu,\Bv)}.
\ee

\vskip2mm
\noindent
\textbf{Step 2.}
Consider the perturbation term in \eqref{eq:GRE.perturb.U.2}
We have
\be\label{eq:eps.R.two.sums}
\bal
\eps_R &= \eps_R(\om) :=
 |\big(\BGni \BU \BG\big)(\Bx,\By)|^s
\\
& \le \left(\sum_{ \diam \Bw \le R/4} + \sum_{\diam \Bw > R/4}\right)
\hesm{ |\BGni(\Bx,\Bw)|^s\,  |\BU(\Bw|^s)\, |\BG(\Bw,\By)|^s }
\\
& \le  \| \BU\|^s \, S_1 +  \eu^{ - \frac{a s}{4}  R } S_2,
\eal
\ee
where
$$
\bal
S_1 &= \sum_{\diam \Bw \le R/4} \hesm{ |\BGni(\Bx,\Bw)|^s\,  |\BG(\Bw,\By)|^s }
\\
S_2 &= \sum_{\diam \Bw > R/4} \hesm{ |\BGni(\Bx,\Bw)|^s\,  |\BG(\Bw,\By)|^s }
\eal
$$
and the factor $\eu^{ - \frac{a s}{4}  R }$ is
of course an upper bound on the interaction over the set $\{\Bw: \diam(\Bw)>R/4\}$.
Using the Cauchy--Schwarz inequality and  the estimate \eqref{eq:thm.apriori.EFC.bound}, we obtain
\be\label{eq:esm.eps.R.2}
\bal
S_2 & \le \sum_{\diam \Bw > R/4} \left( \hesm{ |\BGni(\Bx,\Bw)|^{2s} } \right)^{1/2} \,
  \left( \hesm{ |\BG(\Bw,\By)|^{2s} } \right)^{1/2}
\\
& \le \sum_{\Bw \in (\DZ^d)^2 } \left( \hesm{ |\BGni(\Bx,\Bw)|^{2s} } \right)^{1/2} \,
\cdot \frac{\Const}{(1-2s)  |g|^{2s} } ,
\eal
\ee
provided , for $2s<1$, so the above expectations are finite.

By \eqref{eq:EFC.non-int} with $(\Bu,\Bv) = (\Bx,\Bw)$,
\be\label{eq:sum.Bw.supp.U.inf}
\bal
\sum_{\Bw \in (\DZ^d)^2}
\hesm{ |\BGni(\Bx,\Bw)|^{2s} } &\le A  \sum_{\Bw \in (\DZ^d)^2} \eu^{ - m\, \rd(\Bx,\Bw)}
 =: A' < +\infty,
\eal
\ee
since the function $\Bw\mapsto \eu^{ - m\, \rd(\Bx,\Bw)}$ is summable in $(\DZ^d)^2$.
Thus
\be\label{eq:exp.S2.bound}
\eu^{ - \frac{a s}{4}  R} S_2 \le \frac{\Const}{(1-2s)  |g|^{2s} } \eu^{ - \frac{a s}{4}  R }  .
\ee
Introduce the truncated interaction $\BU_{R}$ of range $R/4$,  with the $2$-body potential
$U_R(r) = \one_{\{r\le R/4\}} U(r)$,  coinciding with $\BU(\Bw)$ on the set $\{\Bw: \diam \Bw > R/4\}$.

The same geometrical argument as in the proof of \eqref{eq:lower.bound.dist.Bx.Bw.finite}, with
$r_0$ replaced now by $R/4$, gives
\be\label{eq:dist.Bx.Bw.2}
\min_{\Bw\in\supp\, \BU_{R/4} } \dist(\Bx, \Bw) \ge \frac{R - \frac{R}{4} }{ 2 } = \frac{3R}{8}.
\ee

Arguing as in the proof of Lemma \ref{lem:exp.dH.diamx.diamy}, we obtain
by \eqref{eq:sum.Bw.supp.U.inf} and \eqref{eq:dist.Bx.Bw.2},
\be\label{eq:sum.Bw.supp.U.2}
\bal
\sum_{\Bw\in\supp\, \BU_{R/4} }
\hesm{ |\BGni(\Bx,\Bw)|^{2s} } &\le C \sum_{r \ge 3R/8 } r^{2d} \eu^{ - m\, r }
\\
&\le C' \eu^{ - c m\, R} .
\eal
\ee
Concluding, we obtain, for $0 < s < 1/3$,
\be\label{eq:esm.eps.R.conclude.2}
\hesm{ \eps_R } \le \Const\, |g|^{-s} \left( \eu^{ - c m\, R}  + \eu^{ - \frac{a s}{4}  R } \right).
\ee
Collecting \eqref{eq:GRE.perturb.U.2}, \eqref{eq:EFC.non-int.2} and \eqref{eq:esm.eps.R.conclude.2},  the assertion follows.
\qedhere


\appendices

\section{Perturbed quadratic recursion}
\label{app:quadratic.dyn}

Fix any $0 < \nu < m/4$.
Let $\beta_k := \eu^{2\nu } M_s^{1/2} \hUps_k$, then the recursion \eqref{rec.hUps}
$$
\hUps_{k+1} \le \half M_s \hUps_k^2 + \half A' L_{k+1}^p \eu^{- \frac{m}{2} L_k}
$$
implies
$$
\eu^{-2\nu } M_s^{-1/2} \beta_{k+1} \le \half \eu^{-4\nu } M_s^{-1}\beta_k^2 + \half A'' L_{k+1}^p \eu^{- \frac{m}{2} L_k},
$$

$$
\beta_{k+1} \le \half \eu^{-2\nu } \beta_k^2 + \half A'' L_{k+1}^p \eu^{- \frac{m}{2} L_k},
$$
with $A'' = A' M_s^{1/2} \eu^{2\nu}$. If $L_0$ (hence, every $L_k$, $k\ge 0$) is large enough, depending on $\frac{m}{2} - 2\nu>0$, then
$$
\beta_{k+1} \le \half \eu^{-2\nu } \beta_k^2 + \half \eu^{-2\nu } \eu^{- \frac{m}{4} L_k}.
$$

To justify this statement,
recall that it follows from the results of the single-particle variants of the MSA and FMM that the decay exponent $m>0$
(or, more precisely, a rigorous lower bound thereupon)
can be chosen in the form $m := c \ln |g|$, for $|g|$ large enough. For example, it suffices to make use of the techniques
from \cite{AM93} or the finite-volume condition from \cite{ASFH01}, applied to single-site subsets of the lattice $\DZ^1$.
For arbitrarily small $\delta>0$ and $L\gg 1$,
$$
\bal
a |g|^s L^b \eu^{ - \frac{m}{2}  L} & \le a |g|^s L^b \eu^{ - c \ln |g| \, L} \le a |g|^s  \eu^{ - (c - \delta) \ln |g| \, L}
\\
&= \frac{a |g|^{s}}{ |g|^{L\delta}}  \eu^{ - (c - 2\delta) \ln |g| \, L}
\le \eu^{ - (c - 2\delta) \ln |g| \, L}  \le \eu^{ - 2\nu \, L},
\eal
$$
with $\nu$ arbitrarily close to $m/4$.

\ble\label{lem:quad.recursion}
Let the sequence $\{L_k, k\ge 0\}$ satisfy the recursion $L_{k+1} = 2L_k +2$.
Consider a sequence of positive numbers $\{\beta_k\}$
with  $\beta_0 < \eu^{-\nu}$,  for some $\nu>0$,  and satisfying
\be\label{eq:recursion?beta.alpha}
\forall \, k\in\DN \quad  \beta_{k+1} \le \half \eu^{-2\nu} \beta^2_k + \half \eu^{-2\nu} \eu^{-2\nu L_k} .
\ee
Then for all $k\ge 1$
\begin{align}
\label{eq:lem.quad.rec.1}
\beta_k &\le \max\big\{ \eu^{ -\nu L_k}, \, \eu^{- \mu L_k} \big\},
\\
\label{eq:lem.quad.rec.2}
\mu &:= \frac{ \nu  + \ln \beta_0^{-1}}{ 1+ \frac{L_0}{2}} .
\end{align}

\ele

\proof
$\bullet$ First, note that if there exists $j\in\DN$ such that
\be\label{eq:beta.le.alpha}
\beta_j \le \eu^{-\nu L_j},
\ee
then by induction, for all $k \ge j$ we have, using $L_k = \half L_{k+1} - 1$,
\be\label{eq:beta.if.never}
\beta_{k+1} \le
\half \eu^{-2\nu} \eu^{-2\nu L_k} + \half \eu^{-2\nu} \eu^{-2\nu L_k} = \eu^{-2\nu} \eu^{-2\nu L_k}  = \eu^{-\nu L_{k+1}}.
\ee

$\bullet$ Next, suppose that \eqref{eq:beta.le.alpha} never occurs, so for all $k$ we have
\be
\label{eq:a.beta.ge}
\beta_{k+1} > \eu^{ -\nu L_k}.
\ee
Then $\half \eu^{-2\nu}\beta^2_k + \half\eu^{-2\nu} \eu^{-\nu L_k} < \eu^{-2\nu}\beta^2_k$, and
by \eqref{eq:recursion?beta.alpha},
$$
\forall\, k\ge 0 \quad \beta_{k+1} < \big( \eu^{-\nu} \beta_k\big)^2.
$$
By induction, with $\mu$ given by \eqref{eq:lem.quad.rec.2} and $2^k = L_k/(L_0+2)$,
for all $k\ge 0$,
\be\label{eq:exp.mu}
 \beta_k \le \big( \eu^{-\nu}\beta_0 \big)^{2^k}
= \big(  \eu^{-\nu} \beta_0 \big)^{ \frac{L_k}{L_0+2} } = \eu^{ -\mu \, L_k },
\ee
with $\mu = -\ln(a^{2}\beta^2_0)/(L_0+2) = \ln(a^{-1}\beta^{-1}_0)/(1 + L_0/2)$.

$\bullet$ Finally, if \eqref{eq:a.beta.ge} holds on a finite integer interval $[[0, j-1]]$,
and then one has \eqref{eq:beta.le.alpha}, the inequality \eqref{eq:exp.mu} is still valid
for $k\in[[0, j-1]]$, while the bounds \eqref{eq:beta.le.alpha}, \eqref{eq:beta.if.never} take over for the remaining values $k \ge j$.
\qedhere

Consequently,
$$
\hUps_k \le \eu^{2\nu} M_s^{1/2} \eu^{- \tmu L_k}, \;\; \tmu := \min(\nu, \mu).
$$

\section{From the EF correlators to the Green functions}

In this section we provide a detailed proof of a fairly general relation between the resolvents
and the EF correlators, which had been used in numerous works on the FMM. The single- or multi-particle nature
of the Hamiltonian at hand is irrelevant, as long as the fractional moments of the EFC can be
effectively assessed in the intended application(s) of the general relation. To stress this,
we temporarily abandon the boldface notation and simply consider a finite-dimensional operator $H$
which need not (but may) be random. Recall that it suffices for our purposes to establish strong dynamical localization
in arbitrarily large but finite domains in the configuration space (with the remaining work to be done
with the help of the Fatou lemma), so we can indeed restrict our analysis to the finite-dimensional
operators.

\ble\label{lem:EFC.to.GF}
\be
\int_I |\BG(\Bx,\By;E)|^s \, dE \le \frac{2 |I|^{1-s}}{1-s} (\BQ(\Bx,\By))^s.
\ee
\proof
In this deterministic statement, the EFs are fixed, and the only relevant variable  is $E$. The GF is a rational function,
and we divide it into the sum of two terms, according to the signs of the numerators:
$$
\BG = \sum_{E_i:\, c_i \ge 0} \frac{ c_i }{ E_i - E} + \sum_{E_i:\, c_i < 0} \frac{ c_i }{ E_i - E}
=: \BG^{(+)} + \BG^{(-)}.
$$
We have
$$
\int_I |\BG(E)|^s \, dE \le \int_I |\BG^{(+)}(E)|^s \, dE + \int_I |\BG^{(-)}(E)|^s \, dE
$$
Both integrals are assessed in the same way, so we focus on the first one.

It is convenient at this point to introduce probabilistic language, for we are going to apply a standard
technique for the probability distribution functions (PDF), and consider the probability space
$(I, \fB_I, \mes_I)$, where $\fB_I$ is the Borel sigma-algebra and $\mes_I := |I|^{-1}$ the normalized
Lebesgue measure in $I\subset\DR$. Further, consider the measurable function $\BG^{(\pm)}:E\mapsto \BG^{(\pm)}(E)$
(i.e., a "random variable" on $(I,\mes_I)$), and let $F_\pm(t)$ be its PDF:
$$
F_\pm(t) = \mes_I\{E:\, |\BG^\pm| \le t\}.
$$
Then
$$
\int_I |\BG^{(\pm)}(E)|^s \, dE = |I|\;  \hesmI{ |\BG^\pm|^s } = |I|\;  \int_0^\infty t^s \, dF_\pm(t)
$$
By the standard formula of integration by parts for the Stiltjes integral (cf., e.g., \cite{Fe66}*{Lemma V.6.1}), for any $s>0$,
$$
\int_0^\infty t^s \, dF_\pm(t) = s \int_0^\infty t^{s-1}\, (1-F_\pm(t))\, dt
$$
where both integrals converge or diverge simultaneously. The goal of the above transformation is to reduce the estimate
to that of the "tail distribution" (in $(I,\mes_I )$) function $t\mapsto 1-F(t) = \mes_I\{\BG^+>t\}$, and it is motivated by the
Boole identity (cf. Proposition \ref{prop:Boole}), applicable to any rational function with simple, real poles and positive expansion coefficients,
$$
f: t \mapsto \sum_{i=1}^{n} \frac{c_i}{t_i - t},
$$
and stating that
$$
\mes \{ \lam:\, |f(\lam)| > t\} = \frac{2\sum_{i=1}^{n} c_i}{t},
$$
hence
$$
\mes_I \{ \lam:\, |f(\lam)| > t\} \le \frac{2\sum_{i=1}^{n} c_i}{|I|t},
$$
Recall that, in fact, $c_i = \psi_i(x)\psi_i(y)$ (we choose the EFs real), so that
$$
\sum_{i:\, c_i\ge 0 } c_i \le \BQ_+(\Bx,\By), \;\;  \sum_{i:\, c_i < 0 } (-c_i )  \le \BQ_-(\Bx,\By),
$$
where $\BQ_\pm(\Bx,\By)$ are components of the EF correlator:
$$
 \BQ_+(\Bx,\By) + \BQ_-(\Bx,\By) = \sum_{i: E_i\in I} |c_i | \le \BQ(\Bx,\By).
$$
Thus, denoting for brevity $\BQ_\pm \equiv \BQ_\pm(\Bx,\By)$, we have
$$
1 - F_\pm(t) \le \min\left( |I|, 2\BQ_\pm t^{-1} \right)
= |I| \one_{[0, 2\BQ_\pm/|I|]}(t) + \frac{2\BQ_\pm}{t} \one_{[2\BQ_\pm/|I|, +\infty)}(t)
$$
and
$$
\bal
\int_0^\infty t^s \, dF_\pm(t)  &\le  s\int_0^\infty t^{s-1}\, (1-F_\pm(t))\, dt.
\\
& \le
s |I| \int_0^{2\BQ/|I|} t^{s-1}\, dt  +2 s\BQ_\pm  \int_{2\BQ/|I|}^\infty t^{s-2} \, dt
\\
&
= |I| \left(\frac{2\BQ_\pm}{|I|}\right)^s +  2s \BQ_\pm  \,\frac{(2\BQ_\pm)^{s-1}}{ |I|^{s-1}(1-s)}
\\
& = \frac{ (2\BQ_\pm)^s |I|^{1-s}}{1-s} .
\eal
$$
Therefore,
$$
\bal
\int_I |\BG(\Bx,\By;E)|^s \, dE &\le \frac{ 2^s |I|^{1-s} }{ 1 - s} \left( \BQ_+(\Bx,\By))^s + \BQ_-(\Bx,\By))^s \right)
\\
& \le \frac{ 2 (\BQ(\Bx,\By))^s |I|^{1-s}}{ 1 - s},
\eal
$$
where the last inequality follows from
$
\frac{\alpha^s + \beta^s}{2} \le \left( \frac{\alpha + \beta}{2} \right)^s, \, s<1.
$
\qedhere

\ele

\section{Boole's identity}
\label{app:Boole}

While there seems to be a consensus that the result stated below was first discovered and proved
by George Boole in 1857, we hesitate to refer to the original work \cite{Boole} as the source of the most
comprehensive proof.
Instead,  we provide a very short (10 lines) and elementary proof given almost a century later by Loomis \cite{Loo46}.
Boole's identity was rediscovered more than once and extended in various ways
in the theory of the Hilbert transform and gave rise to a number
of interesting applications; cf., e.g., \cite{Kolm25,Pol96, SimPZ09, RJLS96, Stein70} and references therein.

\bpr\label{prop:Boole}
Let be given real numbers $\lam_1 < \cdots < \lam_n$
and positive real numbers $c_1, \ldots, c_n$. Then
$$
\forall\, t>0 \quad \mes\left\{ x\in\DR: \; \left|\sum_i \frac{c_i}{ \lam_i - x} \right| > t \right\} = \frac{ 2 \sum_i c_i}{ t} .
$$
\epr

\proof {[Cf. \cite{Loo46}*{Proof of Lemma 1}].}
We assess first the Lebesgue measure of the set $S_+$ where $f(x) := \sum_i  \frac{c_i}{ x -\lam_i } > t$.
Since for all $x \not\in \{\lam_1, \ldots, \lam_n\}$
$$
f'(x) = \sum \frac{-c_i}{ (\lam_i - x)^2} < 0,
$$
there are exactly $n$ roots $\kappa_i$ of the equation $f(x) = t$, and one has $\lam_i < \kappa_i < \lam_{i+1}$,
$\kappa_n > \lam_n$, thus $S_+ = \sqcup_{i=1}^n I_i$, $I_i = (\lam_i, \kappa_i)$, and $\mes S_+ = \sum_i (\kappa_i - \lam_i)$.

Next, multiplying the equation $f(x) := \sum_i  \frac{c_i}{ \lam_i - x} = t$ by $\prod_i (x - \lam_i)$, we see that
$\kappa_i$ are the roots of the polynomial admitting two equivalent representations
$$
 t \prod_i (x - \lam_i)  - \sum_{i=1}^n c_i \prod_{j\ne i} (x - \lam_j) \equiv t \prod_i (x - \kappa_i) .
$$
The identity for the sub-principal coefficients gives
$
t\sum_i \lam_i + \sum_i c_i = t\sum_i \kappa_i,
$
yielding $\mes S_+ = \sum_i (\kappa_i - \lam_i) = t^{-1} \sum_i c_i $.
Similarly, $\mes\,\{x:\, f(x) < -t\} = t^{-1} \sum_i c_i $, so the assertion follows.
\qedhere

\section{The "one-for-all" principle for the fractional moments}

\ble\label{lem:one.for.all}
Let be given real numbers
$
0 < \stwo < t < \sone < 1
$
and a complex-valued random variable $X$ with finite absolute moments $\hesm{|X|^u}$ of all orders $u\in(0, \sone]$.
Then
\be
\bal
\hesm{ |X|^\ttau}  \le
\big( \hesm{ |X|^{\sone}} \big)^{ \frac{\ttau - \stwo}{ \sone - \stwo }}
\cdot \big( \hesm{ |X|^{\stwo}} \big)^{ \frac{\sone - \ttau}{ \sone - \stwo }} .
\eal
\ee

\ele

\proof
First, represent $\ttau$ as a barycenter of $\sone$ and $\stwo$,
$$
\bal
\ttau = \alpha \sone + (1-\alpha) \stwo
= \frac{\sone (\ttau - \stwo)}{ \sone - \stwo} +  \frac{\stwo (\sone -\ttau)}{ \sone - \stwo},
\eal
$$
and define two H\"{o}lder-conjugate exponents,
$$
\qone = \frac{\sone - \stwo}{ \ttau - \stwo}, \;\; \qtwo = \frac{\sone - \stwo}{ \sone - \ttau }.
$$
Now the claim follows directly from the H\"{o}lder inequality:
$$
\bal
\esm{ |X|^\ttau} &= \esm{ |X|^{\alpha \sone} \, |X|^{ (1-\alpha) \stwo } }
\le \esmp{1/\qone}{ |X|^{ \qtwo s \frac{\ttau - \stwo}{\sone - \stwo}}} \cdot
   \esmp{1/\qtwo}{ |X|^{ \qtwo  s \frac{\sone - \ttau}{\sone - \stwo}}}
\\
& = \esmp{1/\qone}{ |X|^{\sone}} \cdot
   \esmp{1/\qtwo}{ |X|^{ \stwo}}
\eal
$$
or, more explicitly,
\be
\bal
\esm{ |X|^\ttau}  \le
\big( \esm{ |X|^{\sone}} \big)^{ \frac{\ttau - \stwo}{ \sone - \stwo }}
\cdot \big( \esm{ |X|^{\stwo}} \big)^{ \frac{\sone - \ttau}{ \sone - \stwo }}
\eal
\ee

\qedhere

\bre
{\rm
A sufficient condition for the convergence of the (absolute) moment of order $u\in(0,1)$ is the
upper bound on the tail probabilities for the r.v. $|X|$:
with $F_{|X|}(t) := \pr{ |X| \le t}$ and some $0 < A < +\infty$,
$$
\int_A^{+\infty} \frac{1}{t^{1-u} } \, \big( 1 - F_{|X|}(t) \big) \, dt < +\infty.
$$
In turn, the above condition is follows from a more explicit bound, often available
in applications: for all sufficiently large $t>0$,
$$
1 - F_{|X|}(t) \le \frac{\Const}{ t^{u^+} }
$$
(the latter means, as usual, an upper bound by $C t^{-(u+\delta)}$ for some $\delta>0$.)
}
\ere

\section*{Acknowledgements} It is a pleasure to thank Michael Aizenman, Simone Warzel, Jeffrey Schenker, G\"{u}nter Stolz, Ivan Veseli\'c and Misha Sodin
for a number of fruitful discussions, related directly or indirectly to the Fractional Moment Method in its various forms.

\end{document}